\documentclass[bibyear]{aa} 

\usepackage{graphicx}
\usepackage{txfonts}
\usepackage{hyperref}
\begin{document}
   \titlerunning{Anomalies in data from the Kepler Legacy project}
  \title{Anomalies in the Kepler Asteroseismic Legacy Project Data \\A re-analysis of 16 Cyg A\&B, KIC8379927 and 6 solar-like stars}
   \author{Ian W. Roxburgh}
   \institute{Astronomy Unit, Queen Mary University of London, 
     Mile End Road, London E1 4NS, UK.
   \email {I.W.Roxburgh@qmul.ac.uk} }
   \date{Received  28 April 2017/ Accepted  }
  \abstract{I compare values of the frequencies, separation ratios, errors and covariance matrices from a new analysis of 9  solar-like stars  with  the Legacy project values reported by Lund et al and, for 16Cyg A\&B and KIC8379927, with values derived by Davies et al.  There is  good agreement  between  my results  
  and Davies's for these 3 stars, but no such  agreement with the Legacy project results.
  My frequencies differ from the  Legacy values, there are inconsistencies in the Legacy frequency covariance matrices which are not positive definite, and the Legacy errors on separation ratios are up to 40 times larger than mine and the values and upper limits derived from the Legacy frequency covariances. There are similar anomalies for 6 other solar-like stars: frequencies and separation ratio errors disagree  and 2 have non positive definite covariance matrices. There are inconsistencies in the covariance matrices of 27 the  66 stars in the full Legacy set and problems with the ratio errors for the vast majority of these stars}
      \maketitle

\section{Introduction}

The Kepler Asteroseismic Legacy Project (Lund et al, 2017)  analysed 66 Kepler main sequence  targets providing frequencies, separation ratios, error estimates and  covariance matrices.
From the outset of this project I queried the data 
(cf Roxburgh 2015, 2016) so I developed my own mode fitting routine, applied this to the Legacy power spectra for 9 solar-like stars, and here compare my results with the Legacy project's  latest (robust) values.
 
In sections 3 to 7 I compare my results for 3 Kepler targets, 16 CygA\&B and KIC8379927, with  the Legacy values and results from independent analyses by   
 Davies et al (2015a,b, 2016), using Davies' power spectra. My results agree well with those of Davies et al, but do not agree with the Legacy project values. 


The Legacy frequencies are different and the error estimates on separation ratios are up to a factor 40 larger and exceed upper limits derived from covariance matrices by a similar factor. 
The covariance matrices are inconsistent as they have negative eigenvalues and are therefore not positive semi-definite as they should be,  giving negative $\chi^2$ when comparing frequency sets,  


In section 8  I compare Legacy and my  results for a further 6 solar-like Legacy stars; 2  have non positive definite covariance matrices, none give good agreement  on frequencies or separation errors.  In section 9  I inspect the covariance  matrices and errors on separation ratios for all 66 Legacy targets and find  similar anomalies.  Something is amiss with the Legacy data. 

The  differences between the Legacy results and those of Roxburgh and  Davies are clearly shown in Fig 1, which  compares the different frequency sets for 16CygB for modes with heights greater than the background ($S/N>1$) - which are less sensitive to background modelling and misidentification of noise for signal than is the case modes with $S/N<1$. I also gives the  $\chi^2$ of the fits using the different error estimates.  The bottom panel compares errors on the separation ratios $ r_{02}$ from all 4 analyses.  The agreement between Roxburgh and Davies is up to $35$ times better than between the Roxburgh and Legacy values. 


\newpage

 \begin{figure}[t]
\begin{center} 
   \includegraphics[width=8.1cm,height=8.2cm]{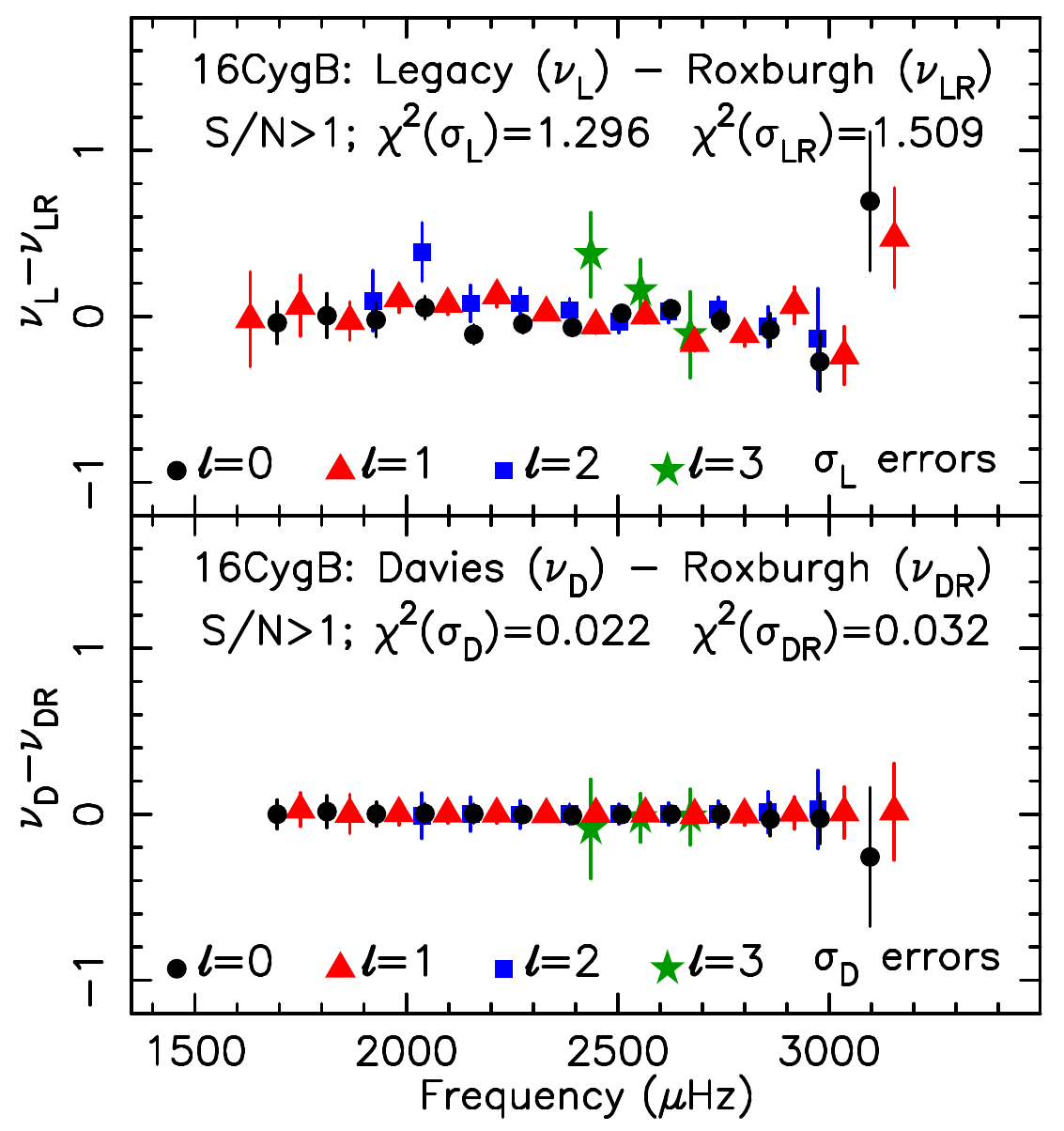}
   \vskip 3pt
\hskip 2pt  \includegraphics[width=7.9cm,height=4.2cm]{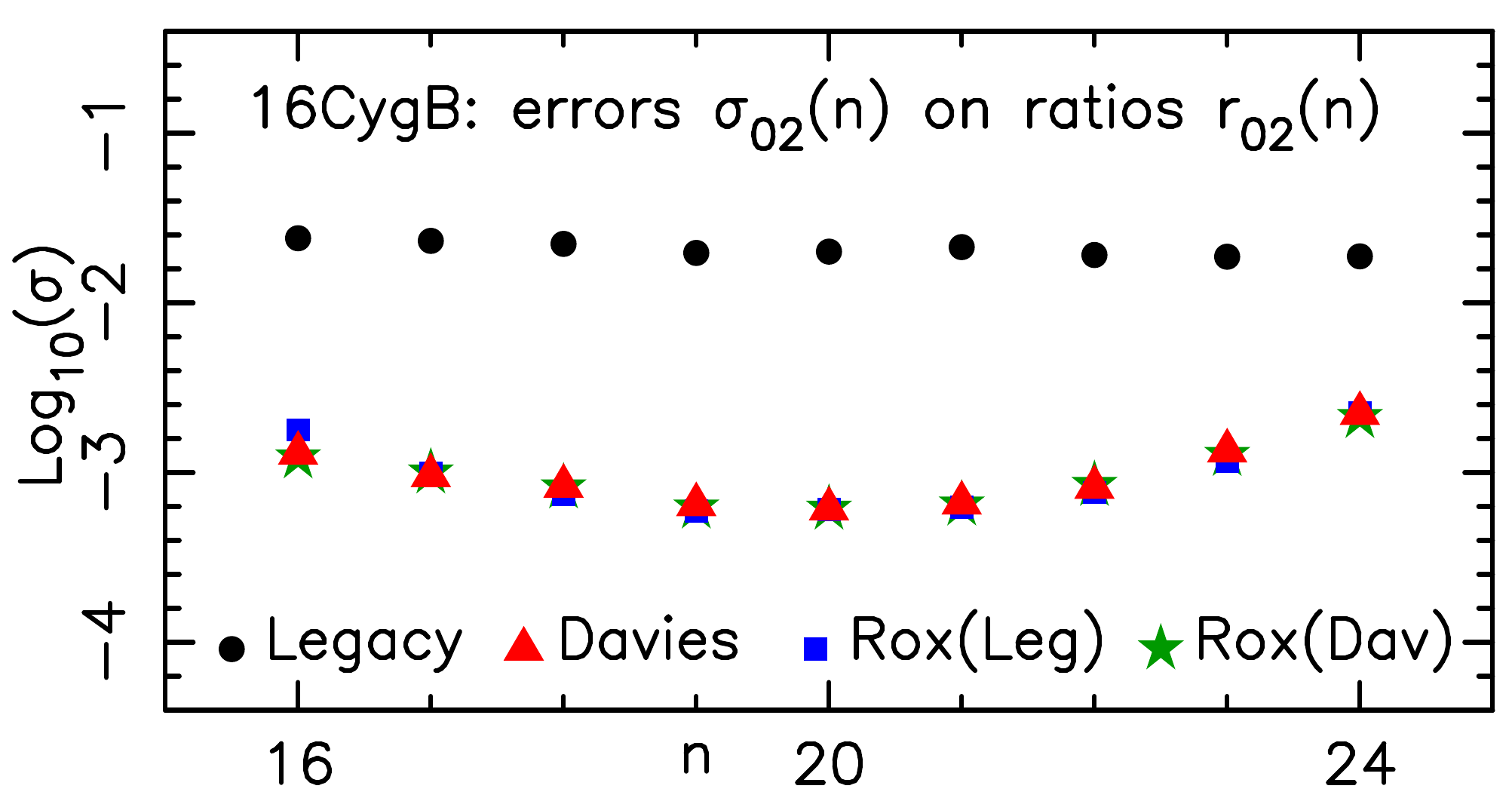}
   \vskip-2pt
   \caption{16CygB: Top 2 panels: frequency differences Legacy-Roxburgh, Davies-Roxburgh and $\chi^2$  of fits;  bottom panel: error estimates $\sigma_{02}$ on the frequency separation ratios $r_{02}$, Legacy, Davies, Roxburgh.}
   \end{center} 
  \vskip-50pt
  \end{figure}
 
\newpage

\section{Roxburgh's mode fitting algorithm}
My mode fitting algorithm searches for a minimum in the negative log likelihood (cf Toutain \& Appourchaux 1994) of a global fit of mode power + background to a section of the power spectrum that extends $\sim 300\mu$Hz beyond both ends of
 the range of frequencies to be fitted,  with unconstrained parameters $X_k$: 
frequencies $\nu_{n,\ell}$;  mode heights $h_n$ and widths $w_n$ of the  $\ell=0$ modes;
 mode height ratios $h_{10}, h_{20}, h_{30}$ of modes $\ell=1,2,3$ to the heights of modes with $\ell=0$ (with the geometrical constraint 
$1+h_{20} =h_{10}+h_{30}$), the same for all modes; rotational splitting  $\nu_{\Omega}$ and inclination $i$ (the same for all modes); and 4 parameters of a Harvey-like model of the background  ($A/[1+B\nu^c] +D$).
The heights and widths of the $\ell=1,2,3$ modes are determined by (linear) interpolation in the values for
 the $\ell=0$ modes at the respective frequencies and, for mode heights, then multiplied by the mode height ratios.
 The  modes are fitted with symmetric rotationally split Lorentzians.
 The covariance matrix is the inverse of the Hessian $H(i,j)=\partial^2 MLE/\partial  X_i\partial X_j$ and the errors on the $X_k$ are given as
 $\sigma_k=[ H^{-1}(k,k)]^{1/2}$.\\
 
\noindent {\bf Power spectra}

\noindent  For comparison with Davies's results I used their power spectra kindly supplied to me by Guy Davies, and for comparison with the Legacy results I used the Legacy power spectra taken from the  kasoc web site namely:
 \begin{table} [h]
\setlength{\tabcolsep}{5.0 pt}
\vskip -6pt
\small
\centering
 \begin{tabular}{l c l c c c c c c c c r c c c   } 
Star/KIC&    kasoc power spectrum&  Quarters \\ [0.5ex]
\hline 
\noalign{\smallskip}
16CygA &kplr012069424\_kasoc-wpsd\_slc\_v1.pow&Q6-17.2\\[0.4ex]
16CygB & kplr012069449\_kasoc-wpsd\_slc\_v2.pow &Q6-17.2\\[0.4ex]
8379927& kplr008379927\_kasoc-wpsd\_slc\_v2.pow&Q2-17.2\\[0.4ex]
 \noalign{\smallskip}
9098294 & kplr009098294\_kasoc-wpsd\_slc\_v1.pow& Q5-17.2  \\ [0.4ex]
8760414 & kplr008760414\_kasoc-wpsd\_slc\_v1.pow &  Q5-17.2 \\  [0.4ex]
6603624 & kplr006603624\_kasoc-psd\_slc\_v1.pow~ ~~& Q5-17.2  \\ [0.4ex]
6225718 &kplr006225718\_kasoc-wpsd\_slc\_v1.pow &   Q6-17.2 \\ [ 0.4ex]
6116048 & kplr006106415\_kasoc-wpsd\_slc\_v2.pow ~~~& Q5-17.2    \\  [0.4ex] 
6106415  & kplr006106415\_kasoc-wpsd\_slc\_v2.pow &  Q6-16.3 \\  [0.4ex]
 \hline 
\end{tabular}
\vskip-10pt
 \end{table}
 
 \noindent 

\section{Results for frequencies:16CygA\&B, KIC 8379927}

Tables 1 to 3 gives the $\chi^2$ of the fits of one set of frequencies to another both for all modes and just for modes with mode-height/background=S/N>1 (as determined by my fits).
I used frequency errors in the fits as I encountered severe problems when using Legacy covariance matrices (see section 5 below).

 Table 1 compares the fit of the Legacy frequencies and errors
 ($\nu_L \pm \sigma_L$) to those of Roxburgh ($\nu_{LR} \pm \sigma_{LR}$) (using the Legacy power spectra), 
$\chi^2_L$ is the value using Legacy errors and $\chi^2_{LR}$ using Roxburgh's errors.  $\chi^2_{LSN}$ is the value using Legacy errors but only comparing frequencies with 
S/N >1, and likewise $\chi^2_{LRSN}$. The first row is for the full frequency sets and the second for frequency sets with "misfits" (discussed below) removed.
Table 2 gives the fit of Roxburgh's frequencies $\nu_{DR}$ (using Davies's power spectra) to Davies's frequencies, $\nu_{D}$ and Table 3 compares the Legacy and Davies's values.  

The Roxburgh-Davies  fit for modes with S/N>1 is very good for all 3 stars, much better than that of Davies's or Roxburgh's fits to the Legacy values. 
The  Roxburgh-Davies fit to 16CygB for all frequencies is strongly influenced by the misfit of the  
$\nu_{14,3}$ mode which has 
S/N=0.15 and is unreliable; 
the Roxburgh-Davies fit for 16CygA for modes with S/N <1
 is strongly influenced by the $\nu_{25,0}$ mode which has $S/N=1.08$, if this is excluded  $\chi^2 _{DSN}=0.023,\, 
\chi^2_{DRSN}=0.026$.

The frequency sets obtained from my analysis for both the Legacy and Davies power spectra,
the Legacy and Davies frequencies, and my S/N values, are given in the Appendix.

\begin{table} [t]
\setlength{\tabcolsep}{8.0 pt} 
\caption { $\chi^2$ of fits of Roxburgh ($\nu_{LR}$) to Legacy ($\nu_L$) frequencies }
\vskip -6pt
\small
\centering
 \begin{tabular}{l c c c c c c c c c c r c c c   } 
\hline\hline 
\noalign{\smallskip}
Star&  $\chi^2_L$  &  $\chi^2_{LSN}$& &   $\chi^2_{LR}$ &  $\chi^2_{LRSN}$\\ [0.5ex]
\hline
\noalign{\smallskip}
16 Cyg A   &   0.791 &   0.897  & & 7.077 & 8.590\\[0.5ex]
16 Cyg A$^*$   &   0.717&   0.805  & & 1.540 &   1.427 \\[0.5ex]
\hline
\noalign{\smallskip}
16 Cyg B   &    1.160  &   1.296  &&  5.067&   1.509 \\[0.5ex]
16 Cyg B$^\dag$   &    1.155&   1.296  &&  1.412&   1.509 \\[0.5ex]
\hline
\noalign{\smallskip}
8379927   &    1.121&   0 .427 && 0.776&   0 .499  \\[0.5ex]
 \hline 
\end{tabular}
 \tablefoot{* Misfits $\nu_{12,0}, \nu_{13,2}$ removed~~
$\dag$  Misfit  $\nu_{12,2}$ removed}
 \end{table}

\begin{table} [t]
\setlength{\tabcolsep}{8.0 pt}
\caption   {$\chi^2$ of fits of Roxburgh ($\nu_{DR}$) to Davies ($\nu_D$) frequencies  }
\vskip -6pt
\small
\centering
 \begin{tabular}{l c c c c c c c c c c r c c c   } 
\hline\hline 
\noalign{\smallskip}
Star& $\chi^2_D$  &  $\chi^2_{DSN}$ && $\chi^2_{DR}$ &  $\chi^2_{DRSN}$\\ [0.5ex]
\hline 
\noalign{\smallskip}
16 Cyg A   & 0.141 &   0 .045  && 0.284 &   0 .062 \\[0.5ex]
16 Cyg A$^*$   & 0.127&   0 .023  && 0.263&   0 .026 \\[0.5ex]
\hline
\noalign{\smallskip}
16 Cyg B    & 0.167&   0 .022  & & 4.227  &   0.032 \\[0.5ex]
16 Cyg B$^\dag$    & 0.137 &   0 .022  & & 0.376&   0 .032 \\[0.5ex]
\hline
\noalign{\smallskip}
8379927     & 0.184 &   0 .033 &&  0.506 &   0 .034  \\[0.5ex]
 \hline 
\end{tabular}
\tablefoot{*For S/N>1.08~~~    
 $\dag$ Misfits  $\nu_{12,2}$ and  $\nu_{14,3}$ removed}
 \end{table}
  \begin{table} [t]
\setlength{\tabcolsep}{8.0 pt}
\caption   {$\chi^2$ of fits of  Davies ($\nu_D$) to Legacy ($\nu_L$) frequencies  }
\vskip -6pt
\small
\centering
 \begin{tabular}{l c c c c c c c c c c r c c c   } 
\hline\hline 
\noalign{\smallskip}
Star&    $\chi^2_L$  &  $\chi^2_{LSN}$& &   $\chi^2_D$ &  $\chi^2_{DSN}$\\ [0.5ex]
\hline 
\noalign{\smallskip}
16 Cyg A&     0.882 &   0 .642 & & 1.424 &   0 .873  \\ [0.5ex]
16 Cyg A$^*$ &     0.784 &   0 .642 & & 0.949 &   0 .873  \\ [0.5ex]
\hline
\noalign{\smallskip}
16 Cyg B&   1.617 &   1.496 &  &  1.786 &   1.910  \\[0.5ex]
16 Cyg B$^\dag$ &   1.631 &   1.496&  &  1.742 &   1.910 \\[0.5ex]
\hline
\noalign{\smallskip}
8379927&  0.936&   0 .663 & &   0.586 &   0 .570 \\[0.5ex]
 \hline 
\end{tabular}
\tablefoot{* Misfits $\nu_{12,0}, \nu_{13,2}$ removed~~    
 $\dag$ Misfits  $\nu_{12,2}$ and  $\nu_{14,3}$ removed}
 \vskip-10pt
  \end{table}

 \vskip 2pt
Table 4 compares the rotational  parameters as determined by Davies et al and as determined by Roxburgh's fits to both the Legacy and Davies power spectra; there is very good agreement for all 3 stars, the fits to the Legacy spectra yielding  almost the same values as obtained in fitting the Davies spectra. 
 \begin{table} [h]
\setlength{\tabcolsep}{2.8 pt}
\caption  {Fit for rotation Davies, Roxburgh (Dspec), Roxburgh (Lspec)}
\vskip-8pt
\small
\centering
 \begin{tabular}{l c c c c c c c c c c r c c c   } 
\hline\hline 
\noalign{\smallskip}
 & Davies  &  & RoxD &  & RoxL&         \\ [1ex]
Star/KIC & $\nu_\Omega\sin i$  & $ i $ & $\nu_\Omega\sin i$  &$ i$  & $\nu_\Omega\sin i$  &$ i $  \\ [1ex]
\hline 
\noalign{\smallskip}
16 Cyg A   &   $0.41\hskip-2.3pt\pm\hskip-2.3pt0.01$ & $56\hskip-2.3pt\pm\hskip-2.3pt6$ &~~$0.40\hskip-2.3pt\pm\hskip-2.3pt0.01$  & $56\hskip-2.3pt\pm\hskip-2.3pt4$  & ~~$0.40\hskip-2.3pt\pm\hskip-2.3pt0.03$ & $46\hskip-2.3pt\pm\hskip-2.3pt3$\\[1.0ex]
16 Cyg B   &$0.27\hskip-2.3pt\pm\hskip-2.3pt0.02$  & $~36\hskip-2.3pt\pm\hskip-2.3pt12$&~~$0.27\hskip-2.3pt\pm\hskip-2.3pt0.01$  & $34\hskip-2.3pt\pm\hskip-2.3pt3$ & ~~$0.26\hskip-2.3pt\pm\hskip-2.3pt0.01$& $50\hskip-2.3pt\pm\hskip-2.3pt5$  \\[1.0ex]
8379927   & $1.11\hskip-2.3pt\pm\hskip-2.3pt0.03$ & $63\hskip-2.3pt\pm\hskip-2.3pt6$&~~$1.11\hskip-2.3pt\pm\hskip-2.3pt0.03$  & $66\hskip-2.3pt\pm\hskip-2.3pt5$ & $~~1.11\hskip-2.3pt\pm\hskip-2.3pt0.02 $ & $~~80\hskip-2.3pt\pm\hskip-2.3pt10$ \\[1.0ex]
 \hline 
\end{tabular}
\vskip-10pt
 \end{table}

\section{Fitting low frequency modes} 
\begin{figure}[t]
\begin{center} 
   \includegraphics[width=8.5cm]{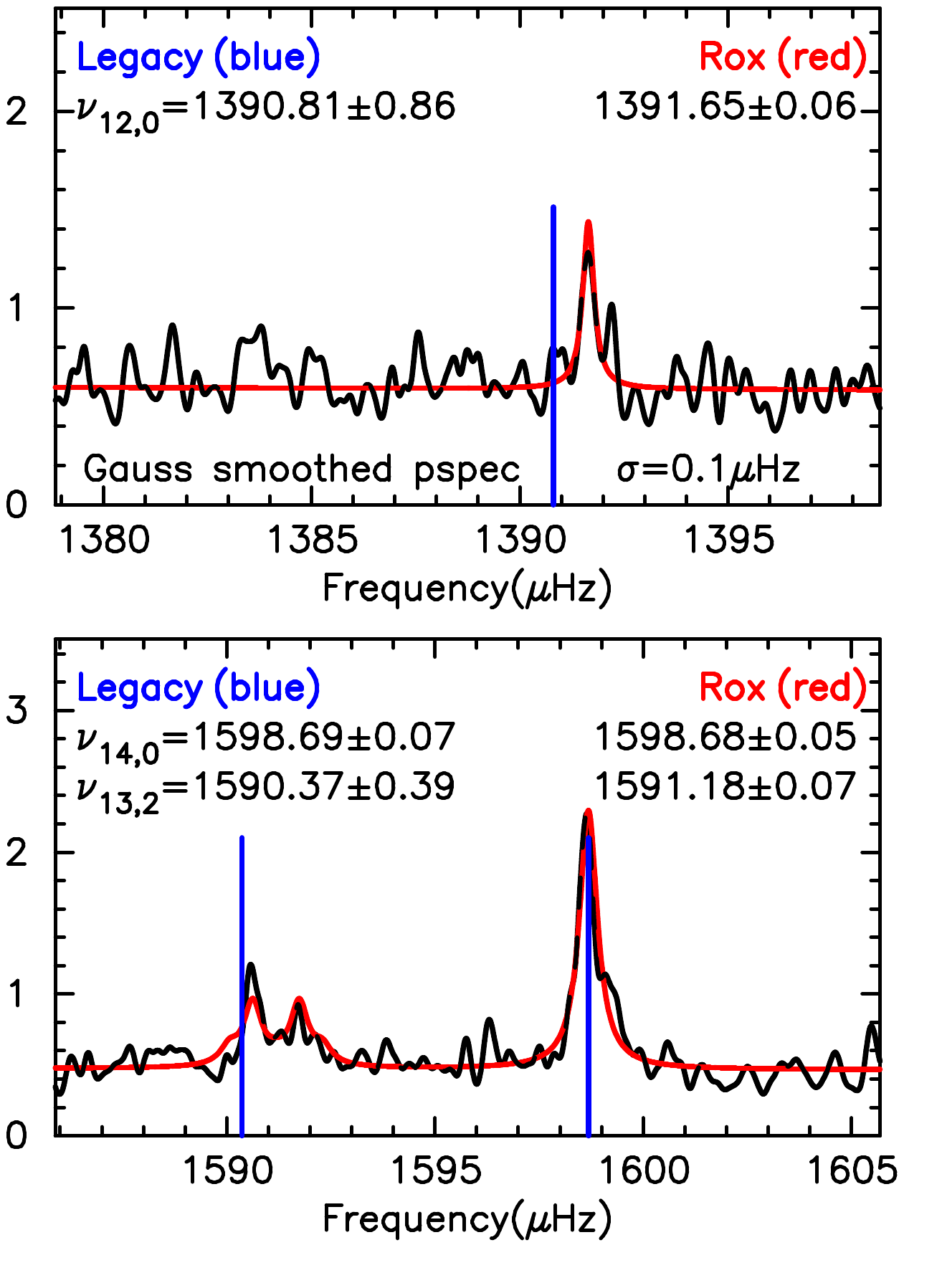}
   \vskip-14pt
   \caption{16CygA:  Legacy and Roxburgh fits  to Legacy power spectrum}
   \end{center} 
   \vskip-15pt
 \end{figure}
\begin{figure}[h]
\begin{center} 
\hskip-12pt
\vskip 4pt
  \includegraphics[width=9cm]{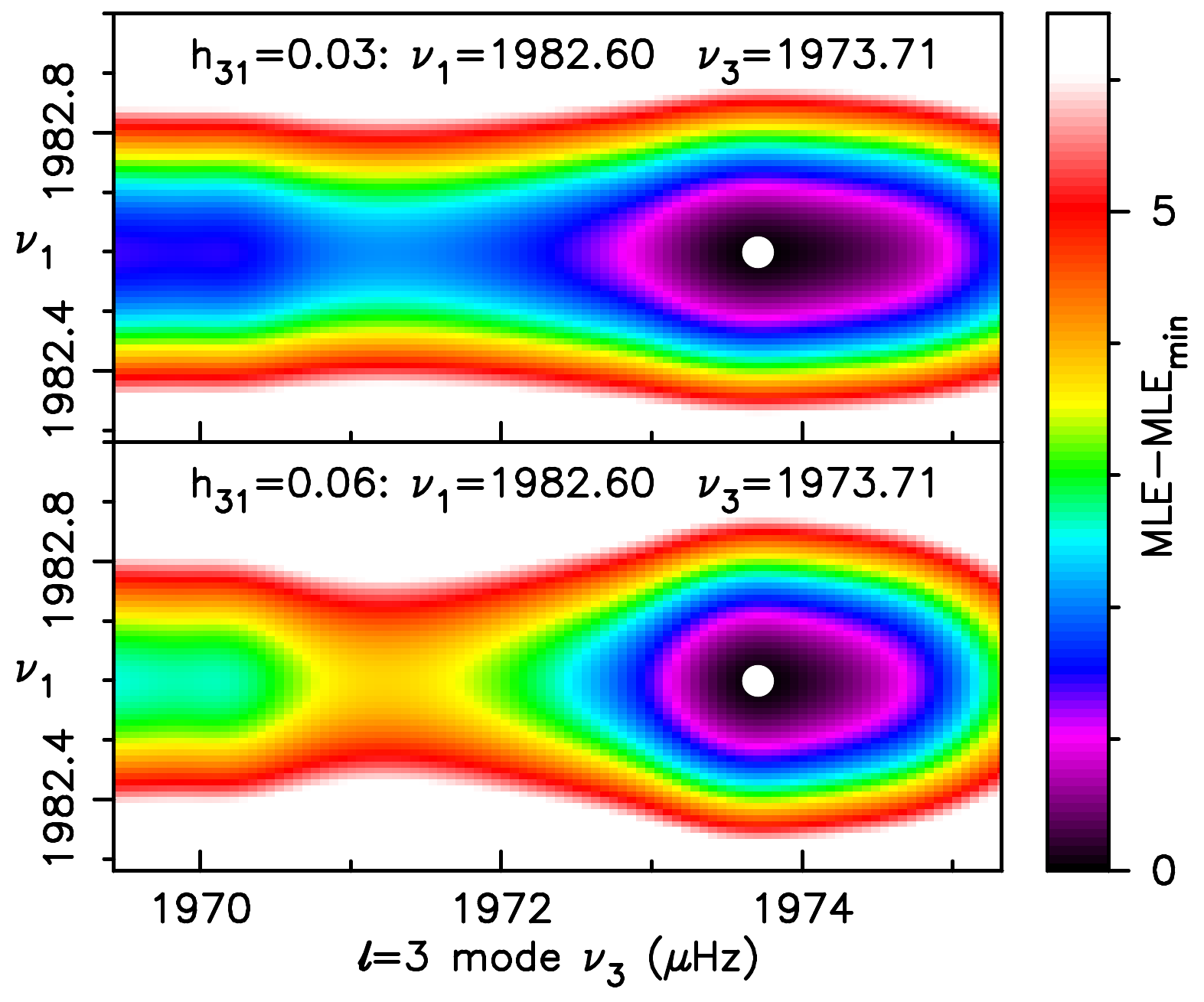}
   \vskip-4pt
   \caption{16CygB:  Quality of free mode fits to Davies  power spectrum}
   \end{center} 
 \vskip-22pt
  \end{figure}
  
As stated above and in the footnotes to the tables there are some problems in fitting some low frequency modes.  For 16CygA Legacy fits (table 1) the problem is
illustrated in Fig 2 which shows the kasoc power spectrum for 16CygA
smoothed by a Gaussian smoother (with FWHM of $0.1\mu$Hz), and overlaid the Roxburgh fit to the full power spectrum and the location of the 
Legacy and Roxburgh frequencies for modes $\nu_{12,0}$ and the pair $\nu_{13.2}, \nu_{14,0}$. The Legacy values for 
$\nu_{12,0}$ and $\nu_{13.2}$ are poor fits and and Roxburgh's error estimates  (from the Hessian of the MLE fit)  
are considerably smaller than the Legacy values. Excluding these 2 modes reduces  $\chi^2_{LR}$ and $\chi^2_{LRSN}$ from 7.077 and 8.590 to
1.540 and 1.427 respectively.
A similar problem exists for the fit to  $\nu_{12,2}$ for 16CygB;  excluding this mode reduces the $\chi^2_{LR}$ of 
 the fit using Roxburgh's errors from 5.067to 1.412.  The S/N values remain unchanged since this mode has S/N<1.

Davies's value of $\nu_{12,2}$ for 16CygB  is also a poor fit to his power spectrum.  $\nu_{14,3 }$ (which has S/N=0.15) differs from my value by $\sim 3\mu$Hz  
so I determined the quality of fits to
  the section of the Davies power spectrum between $1982.6 \pm 29\mu$Hz 
 for a $100^2$ matrix of values of  $\nu_{15,1}, \nu_{14,3 }$ and 10 values of height ratio $h_{31}$ between $0.01$ to $0.1$, with fitting parameters the $\ell=1$ mode height,
 one width  for both $\ell=1$ and $3$ and a constant background; all with $\{\nu_\Omega\sin i, i\}=\{0.27, 34\}$.  
 Fig 3 shows the quality of fits ($MLE-MLE_{min}$) for 2 values of $h_{31}$; the best fits for all $h_{31}$  have $\nu_{14,3}= 1973.71\mu$Hz;  my full fit value is $1973.69\pm0.37\mu$Hz.
 \begin{table} [h]
 \vskip-10pt
\setlength{\tabcolsep}{8.0 pt}
\caption { $\chi^2$ of covariance fits of Roxburgh ($\nu_{LR}$) to Legacy ($\nu_L$) }
\vskip -3pt
\small
\centering
 \begin{tabular}{l c c c c c c c c c c r c c c   } 
\hline\hline 
\noalign{\smallskip}
Star&    $\chi^2_L$  &  $\chi^2_{LSN}$& & $\chi^2_{LR}$ &  $\chi^2_{LRSN}$\\ [1ex]
\hline 
\noalign{\smallskip}
   16CygA &        -3.719 &      0.464 &  &      7.072 &      8.572\\[0.5ex]
   \noalign{\smallskip}
  16CygB &        -0.675 &     -0.432 &  &      5.071 &      1.511\\[0.5ex]
 \noalign{\smallskip}
   8379927 &        1.293 &      0.211 &  &      0.797 &      0.504\\[0.5ex]
  \hline
 \hline 
\end{tabular}
\vskip 10pt
\setlength{\tabcolsep}{8.0 pt}
\caption   {$\chi^2$ of covariance fits of  Davies ($\nu_D$) to Legacy ($\nu_L$)  }
\vskip-3pt
\small
\centering
 \begin{tabular}{l c c c c c c c c c c r c c c   } 
\hline\hline 
\noalign{\smallskip}
Star&    $\chi^2_L$  &  $\chi^2_{LSN}$& & $\chi^2_D$ &  $\chi^2_{DSN}$\\ [1ex]
\hline 
\noalign{\smallskip}
  16CygA                   &      -0.564 &     0.945 &  &    1.736 &    0.925\\[0.5ex]
  \noalign{\smallskip}
  16CygB                   &     4.333 &   0.788 &  &     1.963 &    2.299\\[0.5ex]
  \noalign{\smallskip}
   8379927                      &      -0.787 &      0.185 &   &      0.674 &      0.612\\[0.5ex]
  \hline
 \hline 
\end{tabular}
\vskip 10pt
\caption   {$\chi^2$ of covariance  fits of Roxburgh ($\nu_{DR}$) to Davies ($\nu_D$)  }
\vskip-3pt
\small
\centering
 \begin{tabular}{l c c c c c c c c c c r c c c   } 
\hline\hline 
\noalign{\smallskip}
Star&    $\chi^2_D$  &  $\chi^2_{DSN}$& & $\chi^2_{DR}$ &  $\chi^2_{DRSN}$\\ [1ex]
\hline 
\noalign{\smallskip}
16CygA                   &      0.136 &      0.047 &  &      0.293 &      0.061\\[0.5ex]
  \noalign{\smallskip}
 16CygB                     &     0.138&    0,022 &  &     4.271 &      0.032\\[0.5ex]
  \noalign{\smallskip}
    8379927                      &      0.297  &      0.035 &   &      0.503  &      0.034\\[0.5ex]
  \hline
 \hline 
\end{tabular}
\vskip-10pt

 \end{table}
 
\section{Covariance matrices and frequency comparison}
The $ \chi^2$s of the fit of $N$ frequencies incorporating their correlations are given by  $[ D\,  C^{-1} \,D^T]/N$ where $D$ is the vector of frequency differences and $C^{-1}$ the inverse of the frequency covariance matrix $C$.
Tables 5,6,7 give the results of such  fits  for 16CygA\&B and KIC8379927 for both  the full frequency sets  and for  modes with $S/N>1$  using Legacy (L), Davies (D) and Roxburgh (R) inverse covariance matrices (determined using the SVD algorithm). Whilst the $\chi^2$ for the Roxburgh-Davies fits  are compatible (and small) and consistent with the values using frequency errors as given in tables 1 to 3, the $\chi^2$s using the Legacy covariance matrices give negative values, which should not be the case since covariance matrices  and their inverses are necessarily positive semi-definite so should always give positive $\chi^2$.  

Since a symmetric matrix $C$ is  positive semi-definite if and only if all its eigenvalues are non-negative, I determined the eigenvalues for the Legacy covariance matrices for all 3 stars. The absolute value of the eigenvalues $w_j$ is given by SVD and the sign from which of  $det(C-w_j U)$ and  $det(C+w_j U)$, is zero, or closest to zero given rounding errors. [U is the unit matrix] 
 All 3 Legacy covariance matrices  have negative eigenvalues, 16CygA having 10, 16CygB 12, KIC8379927 10.
 The Roxburgh and Davies covariance matrices are all positive definite. 

 The stark difference between Legacy and Roxburgh  matrices is illustrated in Fig 4 which displays their inverse covariance matrices for 16CygA  [magnitude=size of points, black +ve, red -ve].  Something is clearly amiss with the Legacy evaluation of the covariance matrices from their MCMC analysis.

\begin{figure}[t]
\begin{center} 
   \includegraphics[width=7.2cm,height=7.2cm]{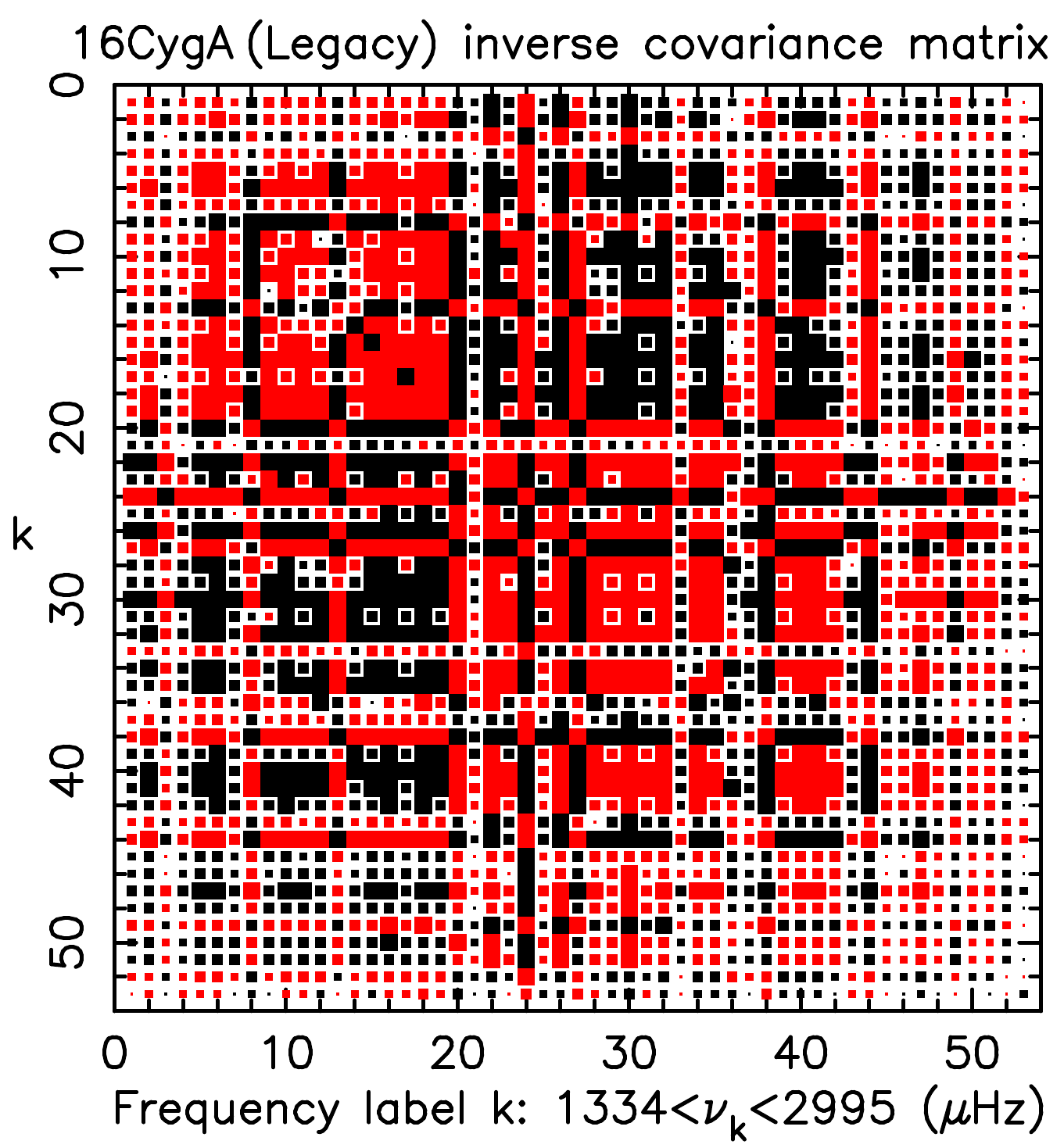}
   \vskip 3pt
     \includegraphics[width=7.2cm,height=7.2cm]{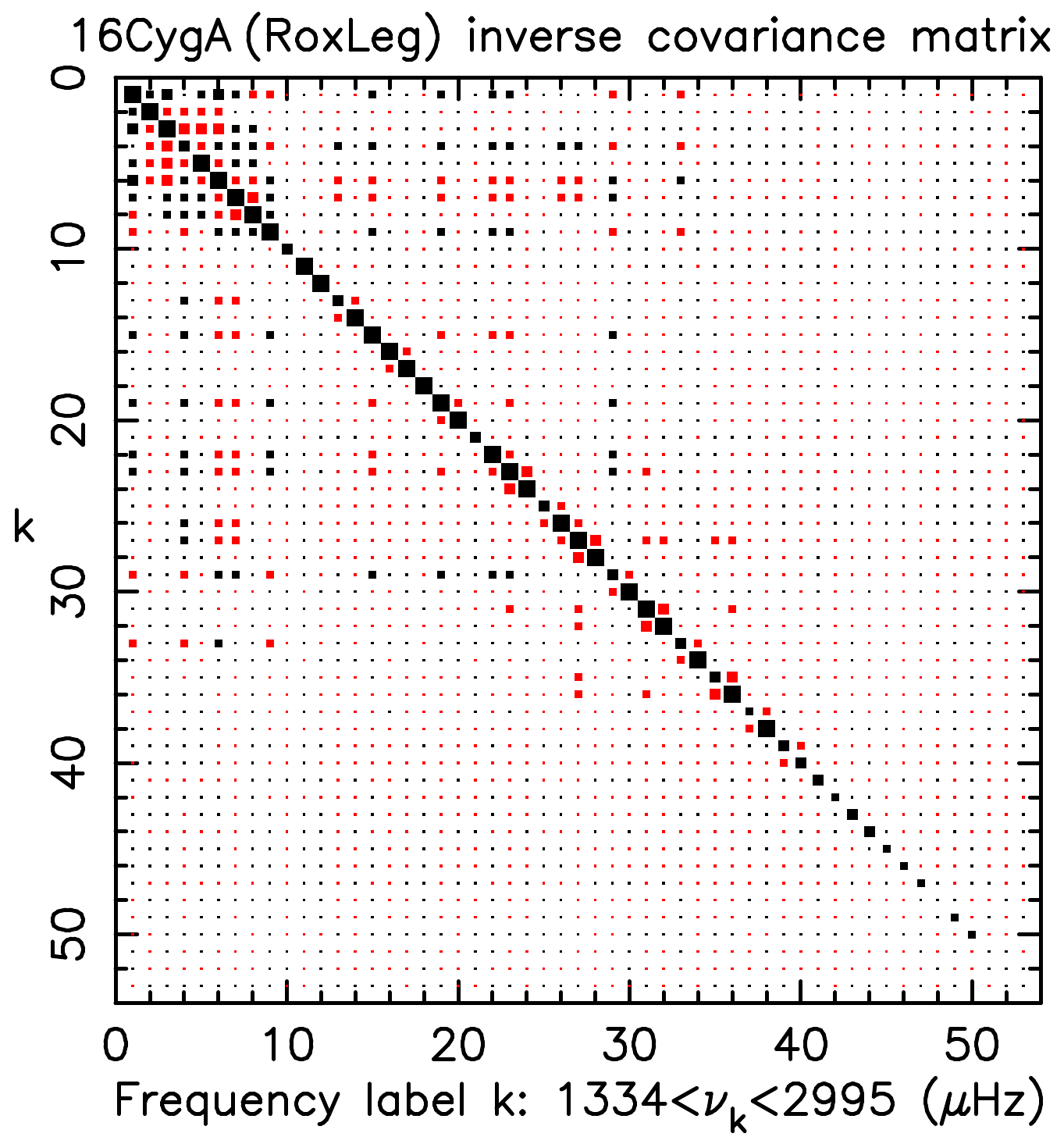}
  \end{center}
  \vskip -10pt
   \caption{16CygA: Inverse covariance matrices. 
   top: Legacy; bottom Roxburgh
[black +ve, red -ve, magnitude = size of points] }
 \end{figure}

\section{Frequency separation ratios}
The ratios of small ($d$) to large ($\Delta$) frequency separations are widely used in model fitting since they are (almost) independent of the structure of the outer layers of a star. These ratios are defined as (Roxburgh \& Vorontsov 2003, 2013, Roxburgh 2005)
$$r_{101}(n)= {d_{101}(n)\over\Delta_n},~~r_{02}(n)= {d_{02}(n)\over\Delta_n}\eqno(1a)$$ 
\vskip -10pt
\noindent where
$$d_{101}(n)= {1\over 8}  \left[ \nu_{n-1,0}-4\nu_{n-1,1}+6\nu_{n,0}-4\nu_{n,1}+\nu_{n+1,0} \right] \eqno(1b)$$
$$d_{02}(n)= \nu_{n,0}-\nu_{n-1,2}, ~~~{\rm and ~~~} \Delta_n=\nu_{n,1}-\nu_{n-1.1} \eqno(1c)$$
\vskip3pt
\noindent The Legacy project and Davies give values of the ratios, errors and ratio covariance matrices for the 3 stars analysed here. They also give values for $r_{010}$ ratios
 but these do not contain any additional information since from 2N
 ($\ell=0,1$)
  frequencies one can only determine N surface layer independent quantities.
 
\begin{figure}[h]
\begin{center} 
   \includegraphics[width=8.6cm]{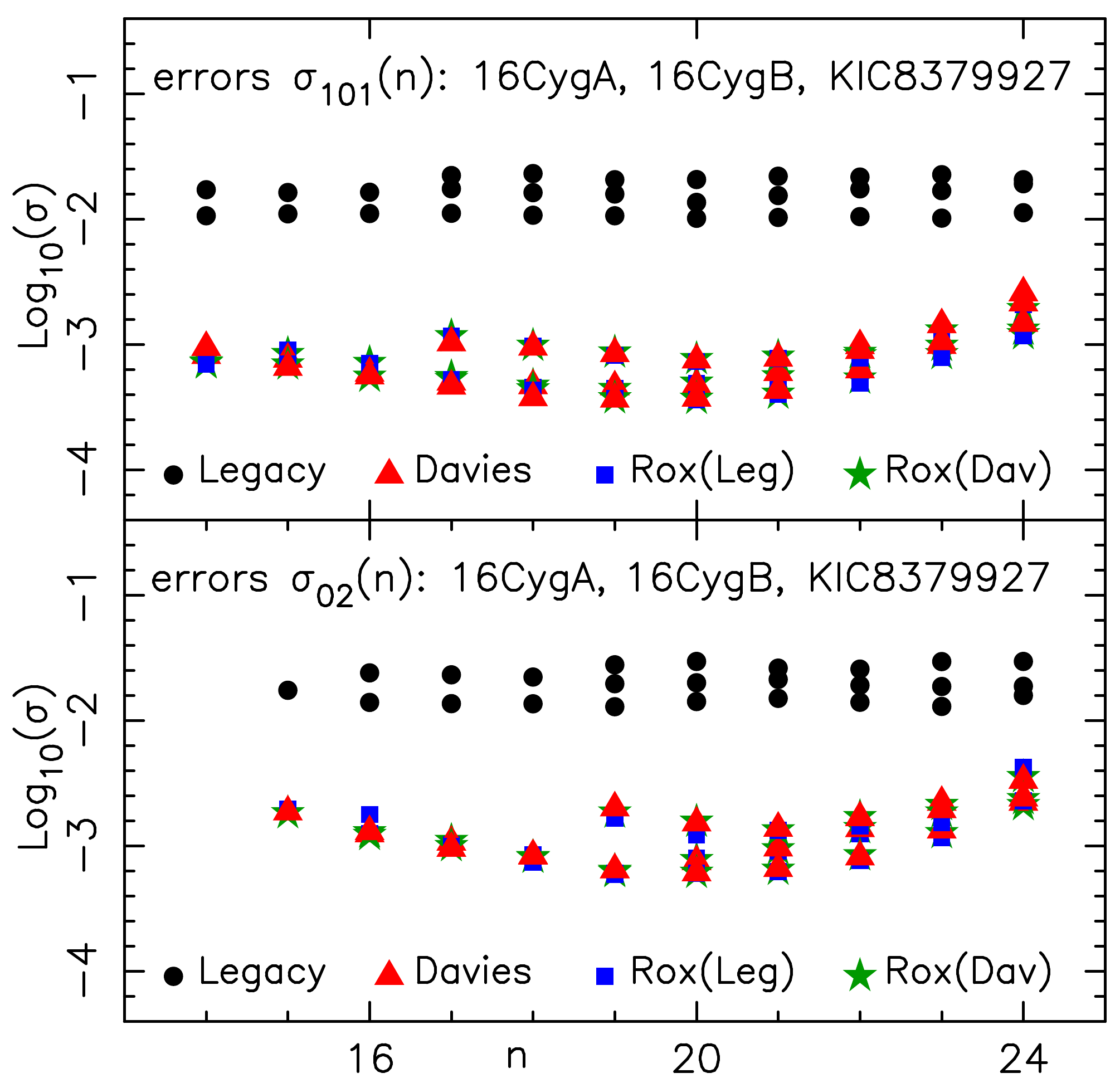}
   \vskip-5pt
   \caption{Top panel: Error estimates $\sigma_{101}$ on ratios  $r_{101}(n)$ from Legacy, Davies, and Roxburgh analyses of 16Cyg A\&B, and KIC 8379927; bottom errors $\sigma_{02}$ on ratios $r_{02}$.}
   \end{center} 
   \vskip-15pt
 \end{figure}

  The values of the ratios  $r_{101}$ and $r_{02}$ as determined by the different analyses are 
 similar but, as shown in Fig 5 the error estimates are wildly different. The top panel shows the 4 determinations of error estimates $\sigma_{101}$ on the ratios $r_{101}$  by Legacy, Davies, and Roxburgh using both the kasoc and Davies power spectra, all limited to modes with S/N>1. The bottom panel shows the error estimates $\sigma_{02}$ on $r_{02}$.
 Davies's and the two Roxburgh values are very close but the Legacy error estimates are very much larger than those of Davies and Roxburgh, by a factor of up to 40.

\section{Error estimates and upper limits for separation ratios  from frequency  covariances}
The covariance of two linear functions $r_n(\nu_j) = \sum  A_{j} \nu_j$, and $r_m(\nu_k)=\,\sum B_{k}  \nu_k$  of  variables $\nu_i$ 
is given by
$$  cov(r_n, r_m)= \sum_j \sum_k  A_{j} \, B_{k} \,cov(\nu_j, \nu_k) \eqno(2) $$
and the error estimate  $\sigma_n$ on $r_n(\nu_k)$ is given by the variance  
$$\sigma_n^2= cov(r_n, r_n)= \sum_j \sum_k  A_{j} \, A_{k} \,cov(\nu_j, \nu_k)= $$
$$ \hskip 70pt\sum_j \sum_k  A_{j} \, A_{k} \,corr_{jk} \,\sigma_j \, \sigma_k \eqno(3)$$
 \noindent where $corr_{jk}$ are the correlations and  $\sigma_i$ the  error estimates on $\nu_i$.
Since $| corr_{jk}| \le 1$  it follows  that  an upper bound on $\sigma_n$ is given by taking $corr_{jk}=+1$ if $A_{j}\,A_{k}>0$ and $-1$ if negative,
hence
$$\sigma_n  \le \sigma_L,~~~\sigma_L^2=\sum_j \sum_{k} |A_{j} \, A_{k}| \,\sigma_j \,\sigma_k \eqno(4)$$

\indent The small separations $d_n$~[both $d_{101}(n)$ and $d_{02}(n)$] are linear functions of $\nu$,~ $d_n=\sum D_k \nu_k$  (cf Eqn 1b, 1c),
but the contribution of the large separation $\Delta_n$ introduces a small non linearity in the ratios. To a good approximation ($1\,{\rm in}\,10^3$ see below) this can be incorporated by expanding around
the average value for the differences $d_0$, large separation $\Delta_0$ and ratios $r_0$ ,which gives 
$$r_n= {d_n\over\Delta_n}={d_n\over\Delta_0} - {d_0\,\Delta_n\over \Delta_0^2} + {d_0\over\Delta_0}=
{1\over\Delta_0} \left[ \sum_k D_{k} \,\nu_k  - r_0\, \Delta_n \right] + r_0 \eqno(5)
$$
which is a linear function of $\nu_k$ as $\Delta_n$ is a linear function of $\nu_k$.  
\eject

The constant term $r_0$ makes no contribution to the covariances so,  with the $A_k$  defined through Eqn 5 (given below),
 the error $\sigma_n$ on $r_n$ is given by Eqn  3 and the upper limit by Eqn 4.

\vskip 15pt
\noindent {\bf Coefficients \boldmath{$A_k$} for the errors \boldmath{$\sigma_{02}(n), \sigma_{101}$} on  \boldmath{$r_{02}(n), r_{101}(n)$}   }
\vskip 4pt
\noindent For $r_0=r_{02}(n)$, $\Delta_0=\nu_{n,1}-\nu_{n-1,1}$
$$\{\nu_k\}=\{\nu_1,~\nu_2, ~\nu_3,~\nu_4 \}= \{\nu_{n-1,1},~\nu_{n-1,2}, ~\nu_{n,0},~ \nu_{n,1}\}$$
$$\{A_k\} =  \left\{{r_0\over \Delta_0}, ~ -{1\over\Delta_0},  ~{1\over \Delta_0}, ~- {{r_0\over \Delta_0}}\right\} \eqno(6)$$
For $r_0=r_{101}(n)$, $\Delta_0=\nu_{n,1}-\nu_{n-1,1}$
$$\{\nu_k\}= \{\nu_1, \nu_2, \nu_3, \nu_4~\nu_5\}= \{\nu_{n-1,0}, \nu_{n-1,1},   \nu_{n,0}, \nu_{n,1}, \nu_{n+1,0}\}$$
$$ \{A_k\} =  \left\{ {1\over 8 \Delta_0}, -{{1-2 r_0}\over  2\Delta_0},  {3\over 4\Delta_0},  -{1+2r_0\over 2\Delta_0},  {{1\over 8 \Delta_0}}\right\} \eqno(7)$$
\vskip 12pt
Fig 6 shows the fractional differences between the $\sigma$'s given by Davies's MCMC analysis of 16CygA\&B and KIC 8379927, and the  $\sigma_{Dcov}$ given by Eqns 3, 6 and 7, using Davies's frequencies and frequency covariances; all but two are less than $10^{-3}$. 
  The two are KIC8379927 $\sigma_{02}(n),\, n=14,15$, which have values $-2.5\,10^{-3},\, 4.4\,10^{-3}$, and are derived from modes with S/N<1.

  Fig 7 shows the same comparison but between Legacy $\sigma$'s and the values $\sigma_{Lcov}$ derived using the Legacy frequencies and frequency covariances;  here many of  the differences are huge.
  
As shown in Fig 8 the Legacy $\sigma$'s also exceed the upper limits $\sigma_u$ given by  Eqn 4, whereas Davies's and Roxburgh's values, and the re-derived Legacy values $\sigma_{Lcov}$,
are less than their corresponding upper limits. Something seems to be amiss with the Legacy values.

\begin{figure}[h]
\begin{center} 
\hskip-10pt
   \includegraphics[width=8.8cm,height=5.0cm]{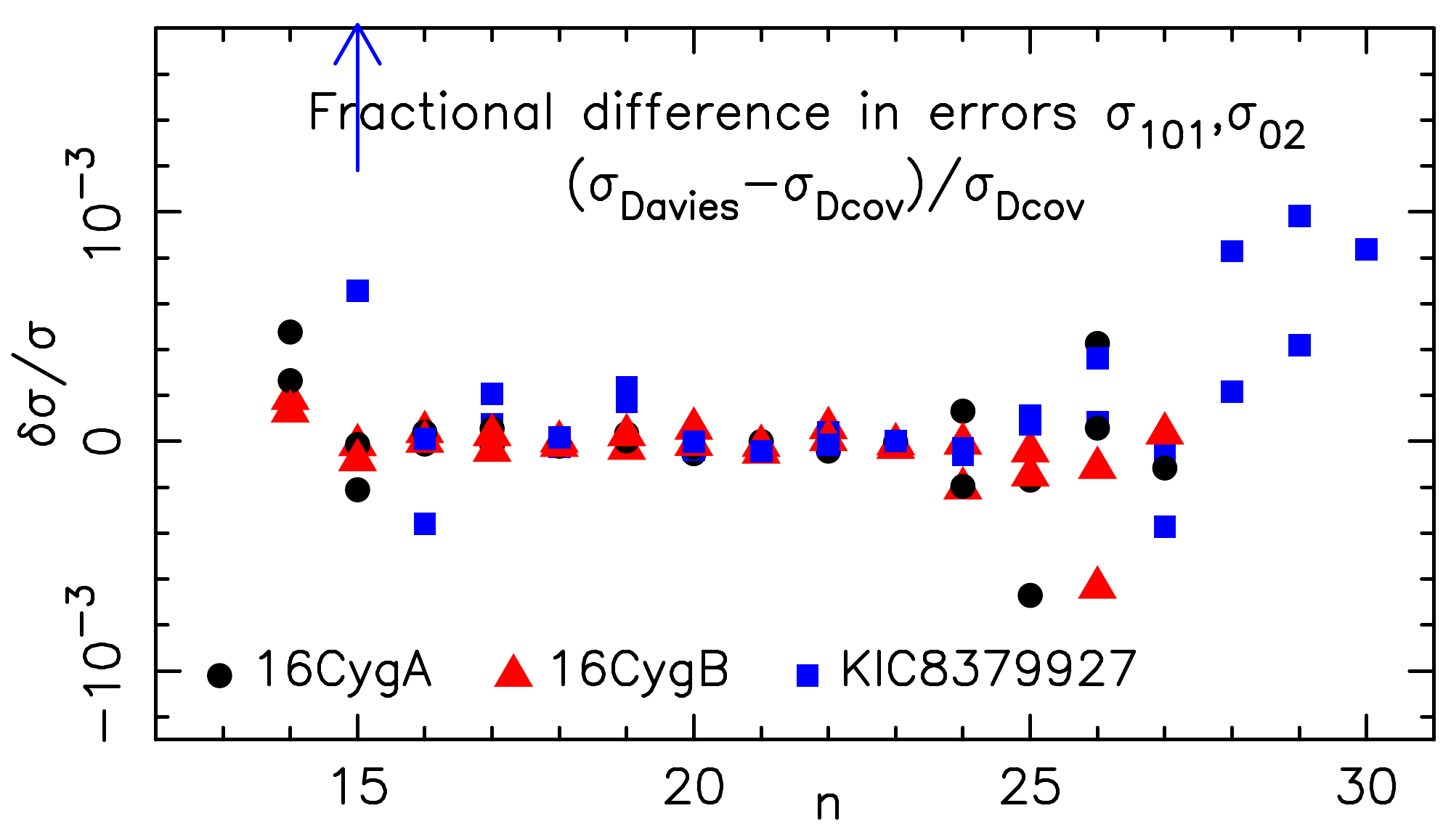}
   \vskip-5pt
   \caption{Fractional difference between Davies's MCMC values for the errors $\sigma_{101}$, $\sigma_{02}$ and the values from Eqns 3, 6, 7.  All but two $<10^{-3}$.}
    \end{center} 
   \vskip-15pt
  \end{figure} 
  
\begin{figure}[t]
\begin{center} 
   \includegraphics[width=8.8cm]{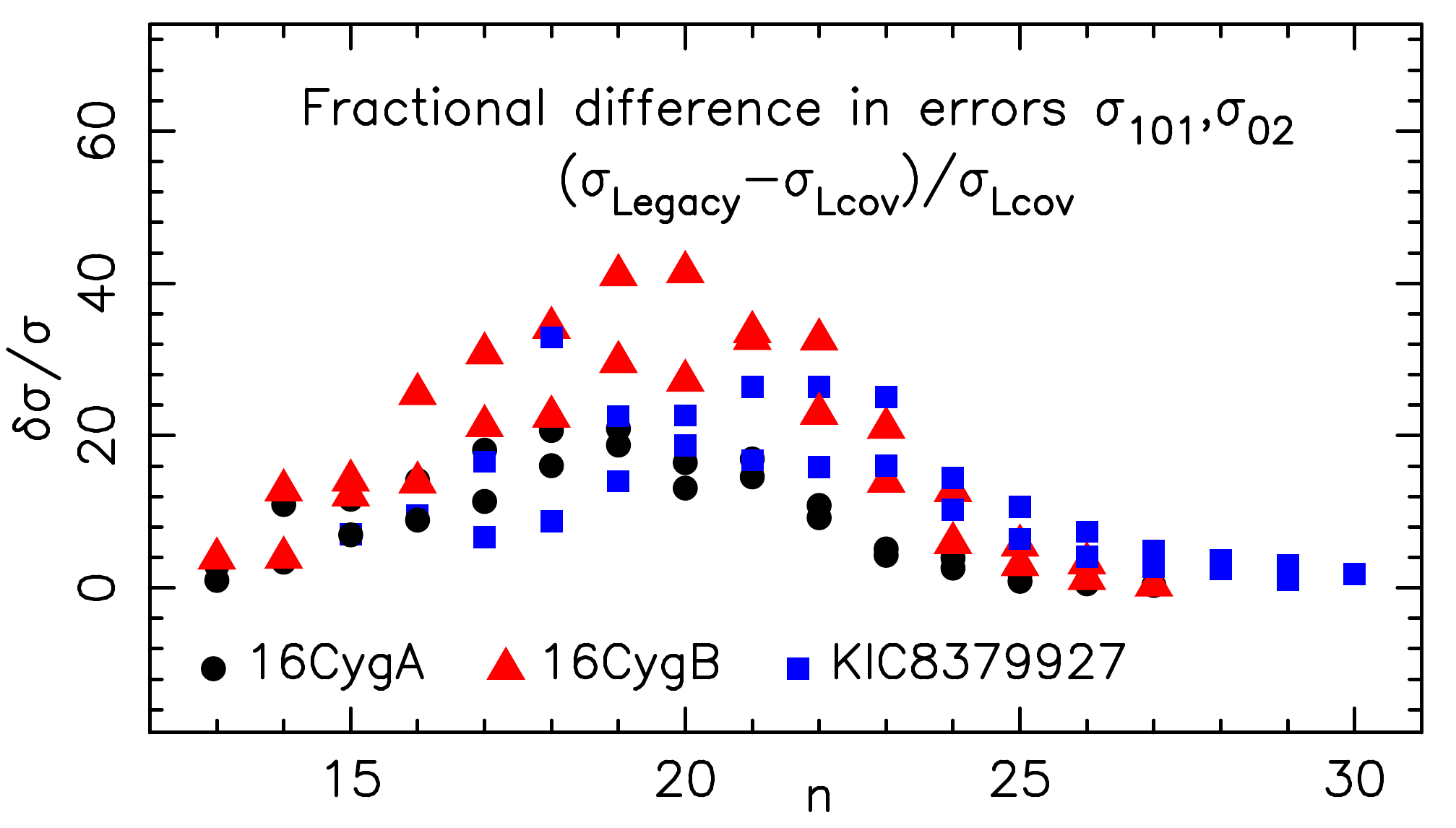}
\vskip-5pt
   \caption{Fractional difference between the Legacy MCMC values for the errors $\sigma_{101}$, $\sigma_{02}$ and the $\sigma_{Lcov}$ values from Eqns 3, 6, 7.}
   \end{center} 
   \vskip-20pt
  \end{figure}
 \begin{figure}[t]
\begin{center} 
   \includegraphics[width=8.8cm]{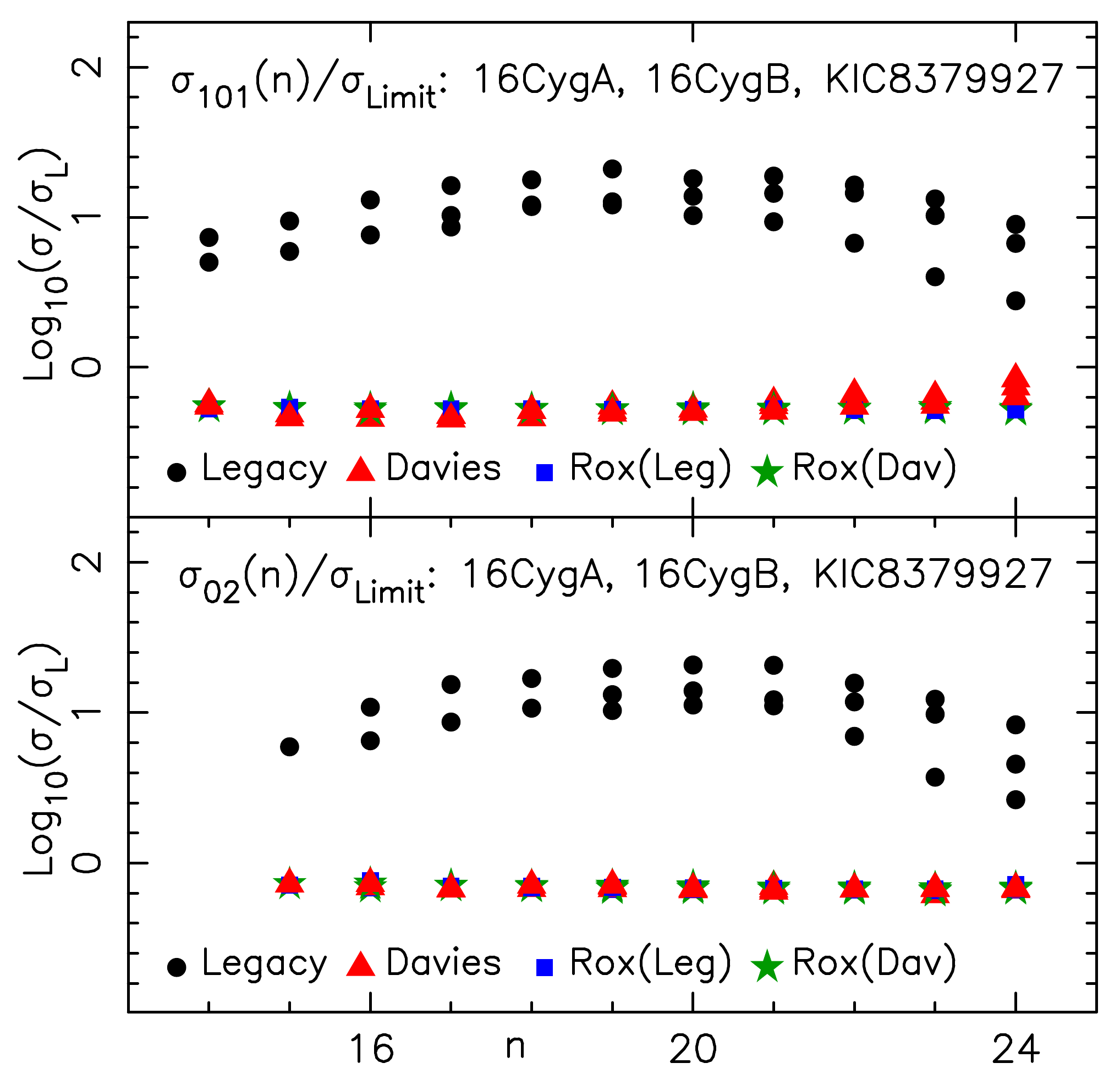}
   \vskip-5pt
   \caption{Ratios of  Legacy error estimates $\sigma_{101}$, $\sigma_{02}$ to the upper limits $\sigma_u$ from Eqn 4 for 16CygA\&B and KIC8379927,
   for modes with $S/N>1$.
   The Legacy values exceed the upper limits by a factor of up to 30. }
   \end{center} 
   \vskip-15pt
  \end{figure}

\section{Comparison of Legacy and Roxburgh results for a further 6 solar-like stars}
 
Having verified that my code gives results in agreement with Davies et al,
I then applied my analysis to the 6 other solar-like stars from the Legacy 
short list of 22 high priority targets which have large separations $\Delta$ in the range $100-120\mu$Hz and $\nu_{max}$ in the range $2138-2470\mu$Hz, namely 
KIC9098294, 8760414, 6603624, 6225718, 6116048, 6106415.  
The fit of the Roxburgh to Legacy frequencies  for KIC6225718 is shown in Fig 9.

Table 8 gives the fits of the Legacy frequencies to Roxburgh's  for all 6 stars using the frequency errors (rows labelled $\sigma$) and the inverse covariance matrices (labelled $cov$) both for all frequencies and for the subset with $S/N>1$. 

\begin{figure}[t]
\begin{center} 
   \includegraphics[width=9cm]{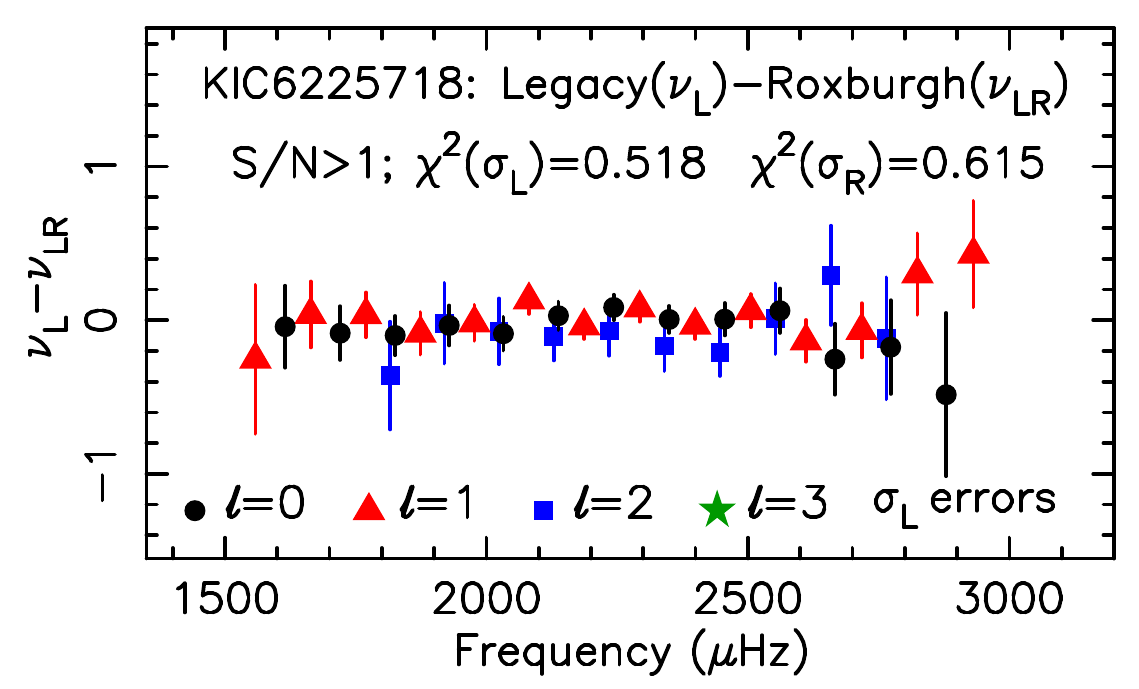}
   \vskip-5pt
   \caption{KIC6225718: Frequency differences, Legacy - Roxburgh }
   \end{center} 
   \vskip-20pt
  \end{figure} 
  
 \begin{table} [t]
 \vskip 10pt
\setlength{\tabcolsep}{6.0 pt}
\caption   {Fit of Roxburgh to Legacy  frequencies: 6 solar-like stars  }
\vskip -6pt
\small
\centering
 \begin{tabular}{l c c c c c c c c c c r c c c   } 
\hline\hline 
\noalign{\smallskip}
Star&  &  $\chi^2_L$  &  $\chi^2_{LSN}$& & $\chi^2_{LR}$ &  $\chi^2_{LRSN}$\\ [1ex]
\hline 
\noalign{\smallskip}
  9098294 & $\sigma$  &      0.535 &      0.505 &  &      0.861 &      0.502\\[0.5ex]
   &    cov &      0.523 &      0.485 &  &      0.814 &      0.487\\[0.5ex]
  \hline
  \noalign{\smallskip}
  8760414 & $\sigma$  &     15.170 &      0.355 &  &     13.462 &      0.380\\[0.5ex]
   &    cov &     26.301 &     -0.041 &  &     16.141 &      0.383\\[0.5ex]
  \hline
  \noalign{\smallskip}
   6603624 & $\sigma$  &      0.762 &      0.226 &  &      1.967 &      0.268\\[0.5ex]
   &    cov &      0.794 &      0.225 &  &      2.001 &      0.268\\[0.5ex]
  \hline
  \noalign{\smallskip}
  6225718 & $\sigma$  &      0.852 &      0.518 &  &      1.141 &      0.615\\[0.5ex]
   &    cov &      1.094 &      0.643 &  &      1.136 &      0.615\\[0.5ex]
  \hline
  \noalign{\smallskip}
  6116048 & $\sigma$  &      0.512 &      0.443 &  &      0.828 &      0.597\\[0.5ex]
   &    cov &      0.176 &      0.486 &  &      0.809 &      0.583\\[0.5ex]
  \hline
  \noalign{\smallskip}
  6106415 & $\sigma$  &      1.479 &      1.100 &  &      1.671 &      1.169\\[0.5ex]
   &    cov &      2.106 &      1.294 &  &      1.592 &      1.166\\[0.5ex]
  \hline
  \hline 
\end{tabular}
 \end{table}
\begin{figure}[h]
\begin{center} 
   \includegraphics[width=8.25cm]{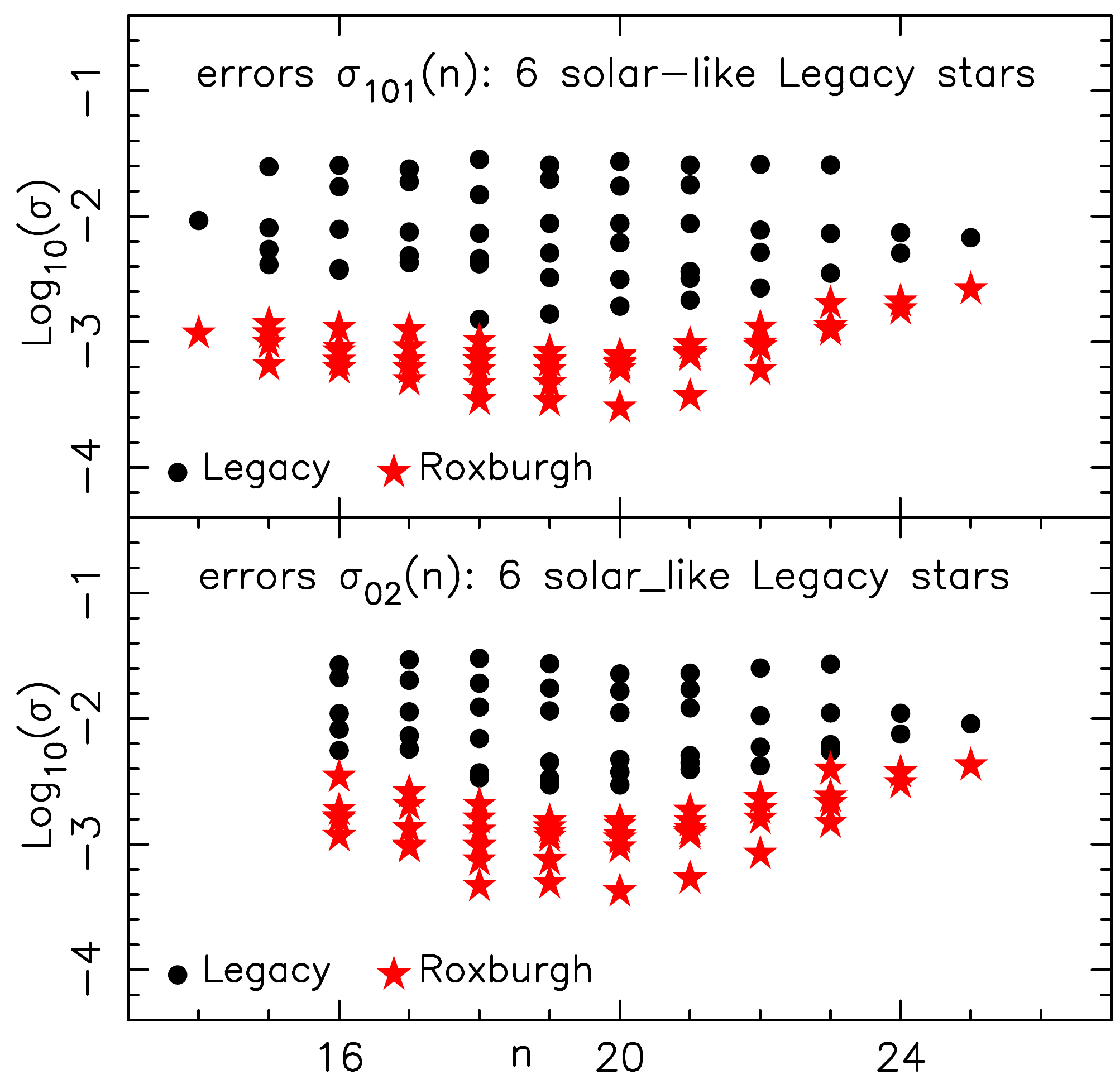}   
   \vskip-5pt
   \caption{The logarithm of the error estimates on the ratios $r_{101}, r_{02}$  for all 6  solar-like stars.
   The Legacy values exceed Roxburgh's values by factors ranging from 2 to 50.}
   \end{center} 
   \vskip-15pt
\end{figure} 

For KIC 9098294, 6603264, there is good agreement between  $\chi^2$ using the Legacy covariance matrices and uncorrelated errors, and reasonable agreement for
 6225718 and 6106415, but the $\chi^2$ are still an order of magnitude larger than the 3 Roxburgh-Davies fits for S/N>1.  The fits for  KIC8760414 and 6116048 are not 
so good:  KIC 8760414 having a negative $\chi^2$ and  KIC 6116048 a factor 3 difference  between values with the Legacy covariance matrix and uncorrelated errors. 
Analysis of the covariance matrices revealed that for the best 4 of the 6 stars  the Legacy covariance matrices had no negative eigenvalues and are therefore positive definite,
whilst the other 2 have negative eigenvalues and are therefore inconsistent. 

Fig 10 plots the Legacy error estimates $\sigma_{101}$ and $\sigma_{02}$ on the separation ratios $r_{101}$ and $r_{02}$ which show a similar behaviour to those of 16CygA\&B and KIC 8379927 in that the Legacy estimates are all larger than Roxburgh's for all 6 stars.  
For  KIC9098294 this is only by a factor $\sim 2$ but for KIC6116048 the Legacy value is up to a factor  $50$ larger then Roxburgh's.  
 
As was the case for 16CygA\&B and KIC8379927, the Legacy ratio errors for all of these 6 stars also exceed the upper limits calculated as described in section 7 above, and likewise new values for errors on the  Legacy ratios calculated using the  Legacy covariance matrices gave  lower values, all of which are less than the corresponding upper limits.

\section{Covariance matrices and  errors on separation ratios  for all 66 Legacy target stars}
The Legacy Project analysed a total of 66 main sequence stars (Lund et al 2017) only 9 of which have been analysed by my code and compared with the Legacy data.
Whilst  this may ultimately be expanded to all the  Legacy targets, I here just examine the Legacy data on all 66 stars to see whether  
their covariance matrices are positive semi-definite or whether they have negative eigevnalues,  and whether they have anomalously large error estimates $\sigma_{101}, \sigma_{02}$ for the separation ratios $r_{101}, r_{02}$.
 
The eigenvalues of all 66 Legacy covariance matrices were determined by the same procedure as applied to 16CygA\&B  and KIC8379927, the absolute magnitudes $w$  from SVD, and the sign from the determinants $||C\pm wU||$.  27 have  covariance matrices with negative eigenvalues and are therefore inconsistent, the remaining 39 stars have positive definite covariance matrices.   
\begin{figure}[t]
\begin{center} 
   \includegraphics[width=8.7cm]{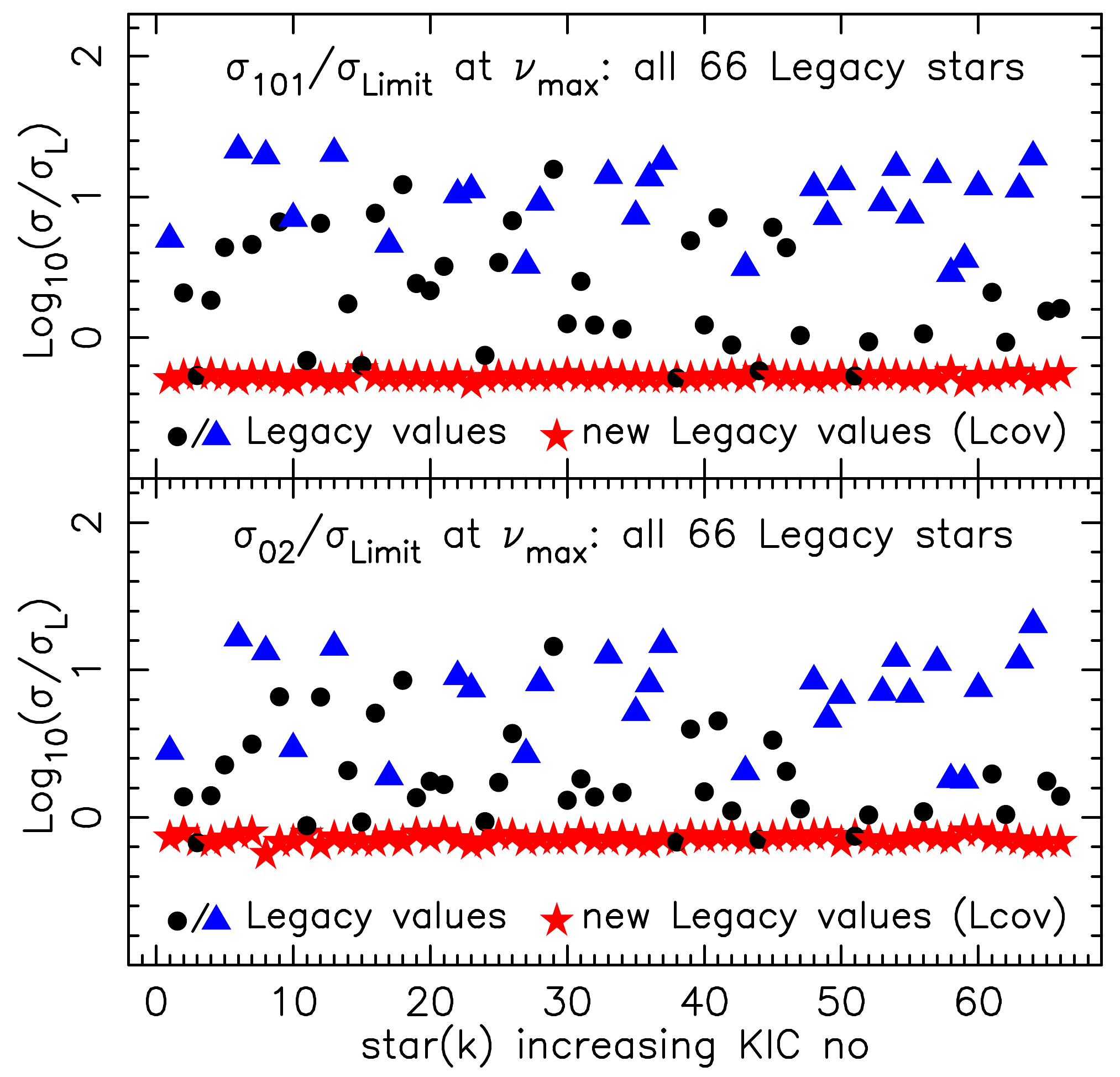}   
   \vskip-5pt
   \caption{The logarithm of the ratio of the error estimates to the upper limits on the ratios $r_{101}, r_{02}$  for all 66  Legacy stars. The blue triangles are stars with inconsistent covariance matrices, the red stars the values re-computed from the covariance matrices. 
   59 of the Legacy values exceed one or both upper limits on $\sigma$.}
   \end{center} 
   \vskip-10pt
  \end{figure} 

Next I compare the Legacy values for the error estimates  $\sigma_{101}, \sigma_{02}$ on the separation ratios $r_{101}, r_{02}$ with the values re-derived from the frequency covariance matrices and the upper limits as determined by Eqns 3,4 6 and 7  in section  7.
 Fig 11 shows the Legacy data error estimates $\sigma_{101}, \sigma_{02}$ divided by the upper limits and the re-derived values divided by the upper limits
 all averaged over 3 values around their $\nu_{max}$. The blue triangles are stars with inconsistent covariance matrices (negative eigenvalues).  
 7 stars have values of $\sigma_{101}$ and $\sigma_{02}$ less than their upper limits  
 all of which have positive definite covariance matrices, of which KIC3427720, 8938364, 9353712 and 10079226   have  Legacy values for ratio  errors  within 2\% of the re-derived values from their covariance matrices. 
 
  I  selected  KIC3427720, the brightest and most solar-like  of the 4 
 to test if, for such a star, the Legacy frequencies agreed  with values obtained on applying my mode fitting algorithm to the power spectrum 
 ( kplr003427720\_kasoc-wpsd\_slc\_v1.pow).
 The details of the fits are given in table 9; as anticipated the $\chi^2$s of the fits using  errors and those using covariance matrices are in good agreement, but the values for modes with S/N>1 are still more than a factor 10 larger than those of the Roxburgh-Davies fits to !6CygA\&B and KIC 8379927.

 \begin{table} [h]
\setlength{\tabcolsep}{6.0 pt}
\caption   {Fit of Roxburgh to Legacy  frequencies: KIC3427720  }
\small
\centering
 \begin{tabular}{l c c c c c c c c c c r c c c   } 
\hline\hline 
\noalign{\smallskip}
Star&  &  $\chi^2_L$  &  $\chi^2_{LSN}$& & $\chi^2_R$ &  $\chi^2_{RSN}$\\ [1ex]
\hline 
\noalign{\smallskip}
  3427720 & $\sigma$  &      0.502 &      0.492 &  &      0.798 &      0.512\\[0.5ex]
                        &    cov &      0.513 &      0.491 &  &      0.810 &      0.512\\[0.5ex]
    \hline
  \hline 
\end{tabular}
\vskip-11pt
 \end{table}

\section{Conclusions and discussion}
{\bf1)} I developed a new mode fitting code different from, and independent of, the codes used by Davies et al and the Legacy project which, when applied to the Davies et al power spectra for 16CygA, 16CygB and KIC 8379927, reproduces the frequencies, separation ratios, errors, rotational parameters and covariance matrices of Davies's analysis  to good accuracy, especially for modes with S/N=heights/background >1 which are least sensitive to  differences in the modelling of the background and the possibility of misidentification of fluctuations in noise as signal. For modes with S/N>1 the $\chi^2$ of the fits of Roxburgh to Davies's frequencies are $\le 0.062$, both  for comparisons using only error estimates and using full covariance matrices ($\le 0.035$ if one mode with S/N=1.08 is excluded).

The same code when applied to the Legacy power spectra for 16CygA, 16CygB and KIC 8379927 does not reproduce the Legacy values. Frequency comparison when using covariance matrices produces anomalous results including negative $\chi^2$; all 3  covariance matrices are inconsistent as they have negative eigenvalues and are therefore not positive semi-definite as any covariance matrix should be. 

 The Legacy errors on separation ratios are up to 40 times larger than my values and exceed values and upper limits derived from the Legacy frequency covariances by a similar factor. 
 
\vskip 3pt
\noindent {\bf2)} I then fitted the power spectra for 6 additional solar-like stars taken from the Legacy high priority list.  Here the agreement is not as bad as for 16CygA\&B and KIC8739927, for modes with S/N>1 the best fit of Roxburgh to Legacy (KIC 6603624) has a $\chi^2 < 0.27$ (still an order of magnitude larger than the Roxburgh-Davies fits), and good agreement between fits using errors and fits using covariance matrices; the worst fit (KIC8760414) gave a negative $\chi^2$ on fitting with the Legacy covariance matrix,  The 4 best fits have positive definitive covariance matrices, the 2 worst fits do not.  For all 6 stars the Legacy error estimates on the separation ratios exceed the values and upper limits  derived using their covariances; KIC9098294 is the one for which the  Legacy values are closest to the values obtained using the frequency  covariances.  

\vskip 3pt
\noindent {\bf3)} Finally I examined all 66 Legacy targets both to test if their covariance matrices were positive semi-definite, and whether or not the errors on separation ratios satisfied their upper limits. The covariance matrices of 27 stars have negative eigenvalues and are therefore inconsistent, 39 have positive  definite covariance matrices. 59 did not satisfy the upper limits on their separation ratio errors, and 4 had ratio errors consistent to within 2\% of the re-derived  values from there (positive definite) covariance matrices.
On fitting the power spectrum of one of these, KIC3427720; my resulting frequencies still did not agree with the Legacy values, all the fits having a $\chi^2\sim 0.5$ whether using Roxburgh or Legacy errors or  covariance matrices.
\vfill\eject
\vskip 3pt
To summarise: results  using my mode fitting code agree with those of Davies et al; results from my code do not agree with the Legacy values; many of the Legacy covariance matrices are inconsistent having negative eigenvalues and therefore are not positive semi-definite;  almost all of the Legacy values for errors on the separation ratios do not agree with values and upper limits derived using the Legacy covariance matrices. 
\vskip 3pt
It is difficult to escape the conclusion that there is something amiss with the Legacy analysis.


\section*{Acknowledgements}
The author thanks Dr G R Davies for supplying and giving permission to use detailed files of the results of his analyses of 16CygA, 16CygB, and KIC8379927;
 Dr M N Lund for  supplying and giving permission to use the updated (robust) results of the Legacy analyses; and a referee for constructive comments.
The author gratefully acknowledges support from the the UK Science and Technology Facilities Council (STFC) under grant  
ST/M000621/1.

\begin{appendix}
 \section{Frequency tables}

The following tables give the frequencies for 16CygA, 16CygB and KIC8379927 as determined by the Legacy project, Roxburgh (Legacy), Davies, and Roxburgh (Davies),
and the values of S/N from my analyses where S/N is defined as the maximum height of a rotationally split mode divided by the local background.
\begin{table*} [h]
\setlength{\tabcolsep}{8.0 pt}
\centerline   {Table A1. 16CygA: Frequencies and errors (in $\mu$Hz) for Legacy, Roxburgh (Legacy), Davies, Roxburgh (Davies)  and S/N values }
\vskip 0.05cm 
\small
\centering
 \begin{tabular}{l c c c c c c c r c c c c c r r c c   } 
\hline\hline 
\noalign{\smallskip}
L&    n  &&$\nu_L$ &  $\sigma_L$&&$\nu_{LR}$ & $\sigma_{LR}$ &~~S/N~&&  $\nu_{D}$ &  $\sigma_{D}$&&$\nu_{DR}$ & $\sigma_{DR}$ &~~S/N~\\ [1ex]
\hline 
\noalign{\smallskip}
     1 &     11 &  &       1334.285 &       1.006 &  &       1334.401 &       0.062 &     0.95 &  &          0.000 &       0.000 &  &          0.000 &       0.000 &     0.00 \\[0.45ex]
     0 &     12 &  &       1390.808 &       0.863 &  &       1391.648 &       0.063 &     1.44 &  &          0.000 &       0.000 &  &          0.000 &       0.000 &     0.00 \\[0.45ex]
     1 &     12 &  &       1437.385 &       0.447 &  &       1437.580 &       0.063 &     1.39 &  &          0.000 &       0.000 &  &          0.000 &       0.000 &     0.00 \\[0.45ex]
     2 &     12 &  &       1487.831 &       0.707 &  &       1487.349 &       0.122 &     0.52 &  &       1488.237 &       0.515 &  &       1488.368 &       0.053 &     0.77 \\[0.45ex]
     0 &     13 &  &       1495.053 &       0.235 &  &       1494.961 &       0.064 &     1.97 &  &       1495.002 &       0.073 &  &       1494.991 &       0.049 &     3.46 \\[0.45ex]
     1 &     13 &  &       1542.060 &       0.140 &  &       1541.952 &       0.054 &     2.45 &  &       1541.922 &       0.065 &  &       1541.906 &       0.048 &     2.07 \\[0.45ex]
     2 &     13 &  &       1590.366 &       0.387 &  &       1591.180 &       0.074 &     1.05 &  &       1591.291 &       0.187 &  &       1591.224 &       0.122 &     0.71 \\[0.45ex]
     0 &     14 &  &       1598.690 &       0.072 &  &       1598.683 &       0.053 &     3.90 &  &       1598.694 &       0.070 &  &       1598.690 &       0.064 &     2.53 \\[0.45ex]
     1 &     14 &  &       1645.140 &       0.109 &  &       1644.996 &       0.099 &     3.76 &  &       1645.063 &       0.086 &  &       1645.046 &       0.088 &     2.67 \\[0.45ex]
     2 &     14 &  &       1693.937 &       0.186 &  &       1694.037 &       0.181 &     1.16 &  &       1694.167 &       0.170 &  &       1694.219 &       0.166 &     1.02 \\[0.45ex]
     0 &     15 &  &       1700.952 &       0.101 &  &       1700.915 &       0.088 &     3.15 &  &       1700.911 &       0.083 &  &       1700.899 &       0.080 &     3.04 \\[0.45ex]
     1 &     15 &  &       1747.199 &       0.085 &  &       1747.181 &       0.081 &     5.23 &  &       1747.149 &       0.076 &  &       1747.150 &       0.080 &     4.67 \\[0.45ex]
     2 &     15 &  &       1795.843 &       0.131 &  &       1795.816 &       0.108 &     2.10 &  &       1795.747 &       0.107 &  &       1795.750 &       0.110 &     1.93 \\[0.45ex]
     0 &     16 &  &       1802.351 &       0.079 &  &       1802.317 &       0.070 &     6.06 &  &       1802.310 &       0.068 &  &       1802.312 &       0.071 &     5.79 \\[0.45ex]
     3 &     15 &  &          0.000 &       0.000 &  &          0.000 &       0.000 &     0.00 &  &       1838.516 &       0.669 &  &       1838.333 &       0.309 &     0.29 \\[0.45ex]
     1 &     16 &  &       1849.009 &       0.056 &  &       1849.016 &       0.056 &     9.83 &  &       1848.976 &       0.053 &  &       1848.977 &       0.056 &     8.29 \\[0.45ex]
     2 &     16 &  &       1898.399 &       0.098 &  &       1898.344 &       0.093 &     3.88 &  &       1898.262 &       0.099 &  &       1898.278 &       0.099 &     3.23 \\[0.45ex]
     0 &     17 &  &       1904.521 &       0.058 &  &       1904.591 &       0.050 &    11.95 &  &       1904.609 &       0.058 &  &       1904.612 &       0.055 &    10.00 \\[0.45ex]
     3 &     16 &  &          0.000 &       0.000 &  &          0.000 &       0.000 &     0.00 &  &       1941.223 &       0.562 &  &       1940.714 &       0.320 &     0.49 \\[0.45ex]
     1 &     17 &  &       1952.008 &       0.050 &  &       1952.027 &       0.049 &    17.66 &  &       1951.996 &       0.049 &  &       1951.997 &       0.050 &    14.11 \\[0.45ex]
     2 &     17 &  &       2001.588 &       0.082 &  &       2001.732 &       0.079 &     6.90 &  &       2001.673 &       0.077 &  &       2001.663 &       0.071 &     5.63 \\[0.45ex]
     0 &     18 &  &       2007.538 &       0.045 &  &       2007.571 &       0.042 &    21.20 &  &       2007.576 &       0.046 &  &       2007.576 &       0.043 &    17.75 \\[0.45ex]
     3 &     17 &  &       2045.851 &       0.368 &  &       2045.876 &       0.229 &     0.88 &  &       2045.976 &       0.365 &  &       2045.912 &       0.195 &     0.84 \\[0.45ex]
     1 &     18 &  &       2055.493 &       0.047 &  &       2055.502 &       0.048 &    29.27 &  &       2055.524 &       0.047 &  &       2055.526 &       0.048 &    23.59 \\[0.45ex]
     2 &     18 &  &       2105.374 &       0.056 &  &       2105.334 &       0.049 &    10.83 &  &       2105.312 &       0.055 &  &       2105.306 &       0.052 &     9.16 \\[0.45ex]
     0 &     19 &  &       2110.949 &       0.041 &  &       2110.900 &       0.041 &    33.95 &  &       2110.909 &       0.039 &  &       2110.914 &       0.040 &    29.91 \\[0.45ex]
     3 &     18 &  &       2150.057 &       0.204 &  &       2149.943 &       0.150 &     1.17 &  &       2149.936 &       0.134 &  &       2149.929 &       0.134 &     1.14 \\[0.45ex]
     1 &     19 &  &       2159.149 &       0.049 &  &       2159.167 &       0.047 &    37.40 &  &       2159.151 &       0.044 &  &       2159.149 &       0.046 &    30.98 \\[0.45ex]
     2 &     19 &  &       2208.928 &       0.072 &  &       2208.956 &       0.069 &    11.60 &  &       2208.900 &       0.064 &  &       2208.894 &       0.064 &     9.97 \\[0.45ex]
     0 &     20 &  &       2214.225 &       0.054 &  &       2214.274 &       0.048 &    32.68 &  &       2214.224 &       0.048 &  &       2214.222 &       0.050 &    28.84 \\[0.45ex]
     3 &     19 &  &       2253.796 &       0.250 &  &       2253.329 &       0.157 &     1.16 &  &       2253.535 &       0.163 &  &       2253.533 &       0.153 &     1.08 \\[0.45ex]
     1 &     20 &  &       2262.562 &       0.051 &  &       2262.552 &       0.051 &    34.77 &  &       2262.537 &       0.048 &  &       2262.534 &       0.049 &    28.91 \\[0.45ex]
     2 &     20 &  &       2312.505 &       0.079 &  &       2312.526 &       0.082 &     9.46 &  &       2312.536 &       0.087 &  &       2312.525 &       0.085 &     7.84 \\[0.45ex]
     0 &     21 &  &       2317.282 &       0.057 &  &       2317.321 &       0.052 &    24.78 &  &       2317.322 &       0.051 &  &       2317.330 &       0.053 &    20.99 \\[0.45ex]
     3 &     20 &  &       2357.497 &       0.227 &  &       2357.226 &       0.200 &     0.91 &  &       2357.392 &       0.189 &  &       2357.341 &       0.198 &     0.79 \\[0.45ex]
     1 &     21 &  &       2366.245 &       0.060 &  &       2366.229 &       0.061 &    24.93 &  &       2366.248 &       0.057 &  &       2366.253 &       0.062 &    20.03 \\[0.45ex]
     2 &     21 &  &       2416.249 &       0.123 &  &       2416.349 &       0.113 &     5.91 &  &       2416.249 &       0.127 &  &       2416.260 &       0.127 &     4.52 \\[0.45ex]
     0 &     22 &  &       2420.937 &       0.080 &  &       2420.959 &       0.079 &    12.50 &  &       2420.897 &       0.081 &  &       2420.920 &       0.084 &     9.35 \\[0.45ex]
     3 &     21 &  &       2461.452 &       0.358 &  &       2461.688 &       0.373 &     0.55 &  &       2462.078 &       0.385 &  &       2461.877 &       0.405 &     0.45 \\[0.45ex]
     1 &     22 &  &       2470.227 &       0.091 &  &       2470.361 &       0.082 &    13.20 &  &       2470.305 &       0.077 &  &       2470.298 &       0.086 &    10.01 \\[0.45ex]
     2 &     22 &  &       2520.734 &       0.199 &  &       2520.618 &       0.174 &     3.15 &  &       2520.459 &       0.212 &  &       2520.475 &       0.191 &     2.51 \\[0.45ex]
     0 &     23 &  &       2524.950 &       0.156 &  &       2525.071 &       0.132 &     5.59 &  &       2525.071 &       0.158 &  &       2525.154 &       0.141 &     4.44 \\[0.45ex]
     3 &     22 &  &          0.000 &       0.000 &  &          0.000 &       0.000 &     0.00 &  &       2566.969 &       0.608 &  &       2567.284 &       0.662 &     0.25 \\[0.45ex]
     1 &     23 &  &       2574.660 &       0.121 &  &       2574.691 &       0.125 &     5.95 &  &       2574.784 &       0.126 &  &       2574.792 &       0.129 &     4.85 \\[0.45ex]
     2 &     23 &  &       2624.636 &       0.362 &  &       2624.975 &       0.369 &     1.34 &  &       2624.322 &       0.324 &  &       2624.331 &       0.334 &     1.21 \\[0.45ex]
     0 &     24 &  &       2628.930 &       0.259 &  &       2629.294 &       0.237 &     2.04 &  &       2629.204 &       0.178 &  &       2629.245 &       0.201 &     1.93 \\[0.45ex]
     3 &     23 &  &          0.000 &       0.000 &  &          0.000 &       0.000 &     0.00 &  &       2669.765 &       1.036 &  &       2668.860 &       1.209 &     0.13 \\[0.45ex]
     1 &     24 &  &       2679.726 &       0.201 &  &       2679.406 &       0.201 &     2.57 &  &       2679.872 &       0.188 &  &       2679.857 &       0.203 &     2.31 \\[0.45ex]
     2 &     24 &  &       2730.024 &       0.756 &  &       2729.839 &       0.546 &     0.80 &  &       2730.233 &       0.886 &  &       2729.550 &       0.716 &     0.68 \\[0.45ex]
     0 &     25 &  &       2733.571 &       0.420 &  &       2734.482 &       0.394 &     1.24 &  &       2733.615 &       0.463 &  &       2734.049 &       0.370 &     1.08 \\[0.45ex]
     1 &     25 &  &       2783.816 &       0.335 &  &       2784.118 &       0.337 &     1.37 &  &       2784.222 &       0.354 &  &       2784.243 &       0.364 &     1.13 \\[0.45ex]
     2 &     25 &  &       2836.088 &       0.798 &  &       2836.291 &       1.692 &     0.31 &  &       2835.339 &       1.147 &  &       2834.364 &       2.878 &     0.24 \\[0.45ex]
     0 &     26 &  &       2840.148 &       0.944 &  &       2838.819 &       1.264 &     0.47 &  &       2838.398 &       0.779 &  &       2838.578 &       0.922 &     0.35 \\[0.45ex]
     1 &     26 &  &       2890.198 &       0.692 &  &       2890.361 &       0.719 &     0.59 &  &       2891.270 &       0.740 &  &       2891.381 &       0.814 &     0.45 \\[0.45ex]
     2 &     26 &  &       2940.393 &       1.103 &  &       2941.252 &       3.200 &     0.18 &  &       2941.479 &       1.538 &  &       2939.644 &       2.527 &     0.15 \\[0.45ex]
     0 &     27 &  &       2944.937 &       0.792 &  &       2945.011 &       1.683 &     0.27 &  &       2945.321 &       1.179 &  &       2946.213 &       1.033 &     0.23 \\[0.45ex]
     1 &     27 &  &       2994.840 &       1.013 &  &       2994.958 &       1.981 &     0.27 &  &       2996.375 &       1.191 &  &       2996.211 &       1.907 &     0.28 \\[0.45ex]
 \hline 
\end{tabular}
\vskip-5pt
 \end{table*}
\newpage

\newpage
\begin{table*} [t]
\setlength{\tabcolsep}{8.0 pt}
\centerline   {Table A2. 16CygB: Frequencies and errors (in $\mu$Hz) for Legacy, Roxburgh (Legacy), Davies, Roxburgh (Davies)  and S/N values }
\vskip 0.05cm 
\small
\centering
 \begin{tabular}{l c c c c c c c r c c c c c r r c c   } 
\hline\hline 
\noalign{\smallskip}
L&    n  &&$\nu_L$ &  $\sigma_L$&&$\nu_{LR}$ & $\sigma_{LR}$ &~~S/N~&&  $\nu_{D}$ &  $\sigma_{D}$&&$\nu_{DR}$ & $\sigma_{DR}$ &~~S/N~\\ [1ex]
\hline 
\noalign{\smallskip}
      1 &     12 &  &       1631.088 &       0.286 &  &       1631.105 &       0.035 &     1.61 &  &          0.000 &       0.000 &  &          0.000 &       0.000 &     0.00 \\[0.45ex]
     2 &     12 &  &       1685.793 &       0.664 &  &       1686.578 &       0.057 &     0.55 &  &       1686.419 &       0.313 &  &       1686.822 &       0.032 &     0.57 \\[0.45ex]
     0 &     13 &  &       1695.023 &       0.126 &  &       1695.061 &       0.063 &     2.25 &  &       1695.069 &       0.087 &  &       1695.069 &       0.053 &     2.71 \\[0.45ex]
     1 &     13 &  &       1749.253 &       0.183 &  &       1749.189 &       0.084 &     1.97 &  &       1749.214 &       0.101 &  &       1749.186 &       0.062 &     2.53 \\[0.45ex]
     2 &     13 &  &       1804.243 &       0.587 &  &       1803.859 &       0.211 &     0.58 &  &       1804.168 &       0.273 &  &       1804.249 &       0.170 &     0.51 \\[0.45ex]
     0 &     14 &  &       1812.444 &       0.133 &  &       1812.440 &       0.068 &     1.89 &  &       1812.428 &       0.097 &  &       1812.412 &       0.078 &     1.59 \\[0.45ex]
     1 &     14 &  &       1866.483 &       0.117 &  &       1866.511 &       0.092 &     2.67 &  &       1866.523 &       0.118 &  &       1866.521 &       0.098 &     2.50 \\[0.45ex]
     2 &     14 &  &       1921.246 &       0.181 &  &       1921.152 &       0.139 &     1.00 &  &       1921.206 &       0.160 &  &       1921.194 &       0.162 &     0.91 \\[0.45ex]
     0 &     15 &  &       1928.886 &       0.103 &  &       1928.908 &       0.076 &     2.86 &  &       1928.901 &       0.072 &  &       1928.899 &       0.070 &     2.53 \\[0.45ex]
     3 &     14 &  &          0.000 &       0.000 &  &          0.000 &       0.000 &     0.00 &  &       1970.959 &       5.137 &  &       1973.695 &       0.375 &     0.15 \\[0.45ex]
     1 &     15 &  &       1982.607 &       0.084 &  &       1982.498 &       0.073 &     4.41 &  &       1982.592 &       0.071 &  &       1982.586 &       0.072 &     4.60 \\[0.45ex]
     2 &     15 &  &       2037.203 &       0.177 &  &       2036.815 &       0.192 &     1.63 &  &       2036.667 &       0.137 &  &       2036.676 &       0.128 &     1.72 \\[0.45ex]
     0 &     16 &  &       2044.357 &       0.069 &  &       2044.305 &       0.067 &     4.44 &  &       2044.278 &       0.060 &  &       2044.273 &       0.058 &     4.97 \\[0.45ex]
     3 &     15 &  &          0.000 &       0.000 &  &          0.000 &       0.000 &     0.00 &  &       2085.370 &       1.498 &  &       2085.478 &       0.353 &     0.24 \\[0.45ex]
     1 &     16 &  &       2098.163 &       0.064 &  &       2098.087 &       0.058 &     7.20 &  &       2098.084 &       0.057 &  &       2098.081 &       0.058 &     7.19 \\[0.45ex]
     2 &     16 &  &       2152.517 &       0.109 &  &       2152.440 &       0.098 &     2.68 &  &       2152.420 &       0.102 &  &       2152.419 &       0.099 &     2.47 \\[0.45ex]
     0 &     17 &  &       2159.503 &       0.058 &  &       2159.612 &       0.061 &     7.40 &  &       2159.581 &       0.057 &  &       2159.577 &       0.059 &     6.35 \\[0.45ex]
     3 &     16 &  &          0.000 &       0.000 &  &          0.000 &       0.000 &     0.00 &  &       2200.579 &       1.224 &  &       2200.453 &       0.354 &     0.37 \\[0.45ex]
     1 &     17 &  &       2214.334 &       0.069 &  &       2214.208 &       0.056 &    12.41 &  &       2214.166 &       0.056 &  &       2214.163 &       0.058 &    11.31 \\[0.45ex]
     2 &     17 &  &       2269.112 &       0.094 &  &       2269.034 &       0.073 &     4.76 &  &       2268.956 &       0.083 &  &       2268.957 &       0.083 &     4.37 \\[0.45ex]
     0 &     18 &  &       2275.949 &       0.054 &  &       2275.994 &       0.049 &    13.13 &  &       2275.948 &       0.048 &  &       2275.949 &       0.047 &    11.44 \\[0.45ex]
     3 &     17 &  &       2318.958 &       0.290 &  &       2318.917 &       0.230 &     0.78 &  &       2319.120 &       0.374 &  &       2319.208 &       0.214 &     0.67 \\[0.45ex]
     1 &     18 &  &       2331.163 &       0.041 &  &       2331.141 &       0.042 &    23.45 &  &       2331.138 &       0.040 &  &       2331.139 &       0.043 &    21.01 \\[0.45ex]
     2 &     18 &  &       2386.252 &       0.070 &  &       2386.214 &       0.057 &     8.82 &  &       2386.263 &       0.061 &  &       2386.262 &       0.060 &     7.66 \\[0.45ex]
     0 &     19 &  &       2392.645 &       0.042 &  &       2392.711 &       0.041 &    26.87 &  &       2392.711 &       0.043 &  &       2392.719 &       0.040 &    22.21 \\[0.45ex]
     3 &     18 &  &       2436.781 &       0.255 &  &       2436.409 &       0.250 &     1.19 &  &       2436.656 &       0.299 &  &       2436.744 &       0.194 &     1.04 \\[0.45ex]
     1 &     19 &  &       2448.181 &       0.048 &  &       2448.237 &       0.041 &    35.07 &  &       2448.253 &       0.041 &  &       2448.251 &       0.042 &    32.20 \\[0.45ex]
     2 &     19 &  &       2503.411 &       0.066 &  &       2503.444 &       0.059 &    11.41 &  &       2503.498 &       0.060 &  &       2503.497 &       0.059 &    10.54 \\[0.45ex]
     0 &     20 &  &       2509.678 &       0.042 &  &       2509.659 &       0.040 &    33.15 &  &       2509.667 &       0.041 &  &       2509.668 &       0.040 &    29.09 \\[0.45ex]
     3 &     19 &  &       2554.181 &       0.188 &  &       2554.026 &       0.125 &     1.41 &  &       2554.146 &       0.147 &  &       2554.167 &       0.157 &     1.25 \\[0.45ex]
     1 &     20 &  &       2565.426 &       0.043 &  &       2565.422 &       0.042 &    40.07 &  &       2565.403 &       0.042 &  &       2565.400 &       0.042 &    37.08 \\[0.45ex]
     2 &     20 &  &       2620.562 &       0.066 &  &       2620.534 &       0.059 &    12.00 &  &       2620.564 &       0.066 &  &       2620.562 &       0.062 &    10.89 \\[0.45ex]
     0 &     21 &  &       2626.458 &       0.050 &  &       2626.413 &       0.045 &    32.90 &  &       2626.397 &       0.045 &  &       2626.397 &       0.044 &    28.60 \\[0.45ex]
     3 &     20 &  &       2671.592 &       0.260 &  &       2671.703 &       0.174 &     1.32 &  &       2671.722 &       0.168 &  &       2671.738 &       0.167 &     1.11 \\[0.45ex]
     1 &     21 &  &       2682.247 &       0.047 &  &       2682.407 &       0.048 &    34.54 &  &       2682.402 &       0.048 &  &       2682.407 &       0.049 &    29.52 \\[0.45ex]
     2 &     21 &  &       2737.707 &       0.075 &  &       2737.666 &       0.073 &     8.74 &  &       2737.744 &       0.079 &  &       2737.743 &       0.080 &     7.02 \\[0.45ex]
     0 &     22 &  &       2743.322 &       0.062 &  &       2743.346 &       0.054 &    20.06 &  &       2743.329 &       0.058 &  &       2743.330 &       0.060 &    14.41 \\[0.45ex]
     3 &     21 &  &       2789.000 &       0.365 &  &       2788.887 &       0.250 &     0.90 &  &       2789.155 &       0.276 &  &       2789.141 &       0.288 &     0.66 \\[0.45ex]
     1 &     22 &  &       2799.613 &       0.072 &  &       2799.721 &       0.062 &    19.92 &  &       2799.734 &       0.063 &  &       2799.737 &       0.065 &    15.06 \\[0.45ex]
     2 &     22 &  &       2855.507 &       0.121 &  &       2855.569 &       0.111 &     4.28 &  &       2855.631 &       0.124 &  &       2855.619 &       0.120 &     3.50 \\[0.45ex]
     0 &     23 &  &       2860.680 &       0.098 &  &       2860.762 &       0.092 &     8.15 &  &       2860.720 &       0.101 &  &       2860.749 &       0.099 &     6.33 \\[0.45ex]
     3 &     22 &  &       2906.905 &       0.490 &  &       2906.922 &       0.391 &     0.44 &  &       2906.865 &       0.435 &  &       2906.862 &       0.445 &     0.34 \\[0.45ex]
     1 &     23 &  &       2917.890 &       0.110 &  &       2917.824 &       0.100 &     8.31 &  &       2917.793 &       0.097 &  &       2917.784 &       0.101 &     6.82 \\[0.45ex]
     2 &     23 &  &       2973.400 &       0.302 &  &       2973.535 &       0.234 &     1.80 &  &       2973.564 &       0.235 &  &       2973.535 &       0.217 &     1.66 \\[0.45ex]
     0 &     24 &  &       2978.180 &       0.175 &  &       2978.454 &       0.157 &     3.09 &  &       2978.504 &       0.151 &  &       2978.529 &       0.145 &     2.79 \\[0.45ex]
     3 &     23 &  &          0.000 &       0.000 &  &          0.000 &       0.000 &     0.00 &  &       3025.061 &       1.128 &  &       3024.682 &       1.203 &     0.16 \\[0.45ex]
     1 &     24 &  &       3035.810 &       0.174 &  &       3036.046 &       0.166 &     3.27 &  &       3036.058 &       0.155 &  &       3036.048 &       0.164 &     2.97 \\[0.45ex]
     2 &     24 &  &       3092.492 &       0.577 &  &       3092.285 &       0.433 &     0.75 &  &       3093.036 &       0.507 &  &       3092.795 &       0.424 &     0.68 \\[0.45ex]
     0 &     25 &  &       3097.170 &       0.419 &  &       3096.476 &       0.403 &     1.27 &  &       3096.850 &       0.419 &  &       3097.107 &       0.372 &     1.10 \\[0.45ex]
     3 &     24 &  &          0.000 &       0.000 &  &          0.000 &       0.000 &     0.00 &  &       3144.035 &       1.415 &  &       3144.275 &       1.288 &     0.07 \\[0.45ex]
     1 &     25 &  &       3154.703 &       0.300 &  &       3154.229 &       0.267 &     1.45 &  &       3154.307 &       0.290 &  &       3154.291 &       0.291 &     1.21 \\[0.45ex]
     2 &     25 &  &       3210.654 &       1.187 &  &       3212.063 &       1.445 &     0.39 &  &       3213.398 &       1.536 &  &       3211.987 &       0.943 &     0.30 \\[0.45ex]
     0 &     26 &  &       3216.451 &       0.482 &  &       3215.846 &       0.533 &     0.66 &  &       3214.925 &       1.040 &  &       3215.900 &       0.697 &     0.48 \\[0.45ex]
     1 &     26 &  &       3273.587 &       0.473 &  &       3273.266 &       0.502 &     0.79 &  &       3273.168 &       0.643 &  &       3273.312 &       0.659 &     0.62 \\[0.45ex]
     2 &     26 &  &       3330.030 &       2.226 &  &       3330.323 &       1.340 &     0.23 &  &       3333.060 &       2.743 &  &       3331.294 &       1.357 &     0.20 \\[0.45ex]
     0 &     27 &  &       3336.009 &       1.060 &  &       3336.187 &       1.516 &     0.39 &  &       3334.219 &       1.903 &  &       3337.847 &       1.054 &     0.33 \\[0.45ex]
     1 &     27 &  &       3391.761 &       1.090 &  &       3393.623 &       0.821 &     0.37 &  &       3393.448 &       0.768 &  &       3393.091 &       0.709 &     0.39 \\[0.45ex]
 \hline 
\end{tabular}
\vskip-5pt
 \end{table*}
\newpage

\newpage
\begin{table*} [t]
\setlength{\tabcolsep}{8.0 pt}
\centerline   {Table A3. KIC8379927: Frequencies and errors (in $\mu$Hz) for Legacy, Roxburgh (Legacy), Davies, Roxburgh (Davies) and S/N values }
\vskip 0.05cm 
\small
\centering
 \begin{tabular}{l c c c c c c c r c c c c c r r c c   } 
\hline\hline 
\noalign{\smallskip}
L&    n  &&$\nu_L$ &  $\sigma_L$&&$\nu_{LR}$ & $\sigma_{LR}$ &~~S/N~&&  $\nu_{D}$ &  $\sigma_{D}$&&$\nu_{DR}$ & $\sigma_{DR}$ &~~S/N~\\ [1ex]
\hline 
\noalign{\smallskip}
     0 &     13 &  &          0.000 &       0.000 &  &          0.000 &       0.000 &     2.25 &  &       1728.138 &       0.435 &  &       1728.025 &       0.141 &     0.72 \\[0.45ex]
     1 &     13 &  &       1783.395 &       0.441 &  &       1783.819 &       0.169 &     0.54 &  &       1783.342 &       0.273 &  &       1783.389 &       0.164 &     0.45 \\[0.45ex]
     0 &     14 &  &       1847.244 &       0.243 &  &       1847.300 &       0.146 &     0.76 &  &       1847.636 &       0.854 &  &       1847.426 &       0.174 &     0.61 \\[0.45ex]
     1 &     14 &  &       1903.592 &       0.282 &  &       1903.637 &       0.131 &     0.68 &  &       1904.672 &       0.846 &  &       1903.864 &       0.191 &     0.47 \\[0.45ex]
     2 &     14 &  &          0.000 &       0.000 &  &          0.000 &       0.000 &     1.00 &  &       1954.857 &       0.725 &  &       1954.580 &       0.678 &     0.15 \\[0.45ex]
     0 &     15 &  &       1967.982 &       0.255 &  &       1967.920 &       0.154 &     0.91 &  &       1968.190 &       0.217 &  &       1968.206 &       0.197 &     0.69 \\[0.45ex]
     1 &     15 &  &       2023.838 &       0.297 &  &       2023.892 &       0.230 &     0.98 &  &       2023.666 &       0.310 &  &       2023.654 &       0.244 &     0.67 \\[0.45ex]
     2 &     15 &  &          0.000 &       0.000 &  &          0.000 &       0.000 &     1.63 &  &       2075.323 &       0.590 &  &       2075.643 &       0.432 &     0.26 \\[0.45ex]
     0 &     16 &  &       2087.937 &       0.198 &  &       2087.844 &       0.152 &     1.27 &  &       2087.990 &       0.156 &  &       2087.960 &       0.144 &     1.05 \\[0.45ex]
     1 &     16 &  &       2143.133 &       0.180 &  &       2143.140 &       0.156 &     1.55 &  &       2143.237 &       0.178 &  &       2143.207 &       0.171 &     1.14 \\[0.45ex]
     2 &     16 &  &       2195.244 &       0.469 &  &       2194.891 &       0.256 &     0.58 &  &       2195.355 &       0.299 &  &       2195.316 &       0.338 &     0.43 \\[0.45ex]
     0 &     17 &  &       2206.506 &       0.150 &  &       2206.692 &       0.126 &     1.97 &  &       2206.657 &       0.126 &  &       2206.635 &       0.120 &     1.78 \\[0.45ex]
     1 &     17 &  &       2261.245 &       0.136 &  &       2261.106 &       0.122 &     2.37 &  &       2261.245 &       0.116 &  &       2261.218 &       0.118 &     1.88 \\[0.45ex]
     2 &     17 &  &       2312.707 &       0.370 &  &       2312.251 &       0.313 &     0.92 &  &       2312.901 &       0.276 &  &       2312.899 &       0.264 &     0.74 \\[0.45ex]
     0 &     18 &  &       2324.439 &       0.106 &  &       2324.451 &       0.112 &     3.16 &  &       2324.322 &       0.111 &  &       2324.307 &       0.109 &     2.80 \\[0.45ex]
     1 &     18 &  &       2379.779 &       0.099 &  &       2379.770 &       0.095 &     3.49 &  &       2379.939 &       0.108 &  &       2379.911 &       0.106 &     2.72 \\[0.45ex]
     2 &     18 &  &       2432.197 &       0.204 &  &       2432.318 &       0.178 &     1.28 &  &       2432.294 &       0.223 &  &       2432.348 &       0.199 &     0.95 \\[0.45ex]
     0 &     19 &  &       2443.152 &       0.101 &  &       2443.116 &       0.096 &     4.34 &  &       2443.150 &       0.103 &  &       2443.131 &       0.096 &     3.61 \\[0.45ex]
     1 &     19 &  &       2499.437 &       0.092 &  &       2499.391 &       0.088 &     5.12 &  &       2499.398 &       0.104 &  &       2499.388 &       0.099 &     3.65 \\[0.45ex]
     2 &     19 &  &       2552.415 &       0.153 &  &       2552.284 &       0.126 &     1.96 &  &       2552.244 &       0.166 &  &       2552.222 &       0.171 &     1.35 \\[0.45ex]
     0 &     20 &  &       2563.543 &       0.086 &  &       2563.596 &       0.077 &     6.75 &  &       2563.605 &       0.082 &  &       2563.587 &       0.081 &     5.00 \\[0.45ex]
     1 &     20 &  &       2619.926 &       0.090 &  &       2619.986 &       0.093 &     6.54 &  &       2619.991 &       0.086 &  &       2619.970 &       0.092 &     4.81 \\[0.45ex]
     2 &     20 &  &       2673.136 &       0.145 &  &       2673.038 &       0.136 &     2.17 &  &       2673.135 &       0.145 &  &       2673.157 &       0.133 &     1.69 \\[0.45ex]
     0 &     21 &  &       2683.948 &       0.099 &  &       2683.962 &       0.088 &     6.92 &  &       2684.018 &       0.097 &  &       2683.995 &       0.092 &     6.02 \\[0.45ex]
     1 &     21 &  &       2740.437 &       0.087 &  &       2740.463 &       0.088 &     6.95 &  &       2740.526 &       0.089 &  &       2740.507 &       0.092 &     5.25 \\[0.45ex]
     2 &     21 &  &       2793.385 &       0.159 &  &       2793.453 &       0.154 &     2.29 &  &       2793.482 &       0.185 &  &       2793.435 &       0.184 &     1.70 \\[0.45ex]
     0 &     22 &  &       2804.566 &       0.087 &  &       2804.494 &       0.087 &     7.05 &  &       2804.495 &       0.093 &  &       2804.480 &       0.096 &     5.06 \\[0.45ex]
     1 &     22 &  &       2860.993 &       0.095 &  &       2861.027 &       0.097 &     6.68 &  &       2861.002 &       0.107 &  &       2860.981 &       0.107 &     4.89 \\[0.45ex]
     2 &     22 &  &       2913.836 &       0.177 &  &       2914.001 &       0.163 &     2.01 &  &       2914.039 &       0.219 &  &       2914.029 &       0.200 &     1.56 \\[0.45ex]
     0 &     23 &  &       2924.530 &       0.094 &  &       2924.483 &       0.092 &     6.06 &  &       2924.469 &       0.112 &  &       2924.447 &       0.107 &     4.42 \\[0.45ex]
     1 &     23 &  &       2981.323 &       0.120 &  &       2981.304 &       0.115 &     5.37 &  &       2981.241 &       0.123 &  &       2981.220 &       0.123 &     4.06 \\[0.45ex]
     2 &     23 &  &       3034.225 &       0.276 &  &       3034.302 &       0.255 &     1.45 &  &       3033.920 &       0.271 &  &       3033.886 &       0.262 &     1.15 \\[0.45ex]
     0 &     24 &  &       3044.841 &       0.130 &  &       3044.850 &       0.131 &     3.48 &  &       3044.862 &       0.129 &  &       3044.853 &       0.129 &     2.92 \\[0.45ex]
     1 &     24 &  &       3102.033 &       0.157 &  &       3101.964 &       0.156 &     3.59 &  &       3102.002 &       0.155 &  &       3101.960 &       0.162 &     2.80 \\[0.45ex]
     2 &     24 &  &       3155.043 &       0.376 &  &       3155.020 &       0.374 &     0.99 &  &       3154.846 &       0.359 &  &       3154.852 &       0.351 &     0.77 \\[0.45ex]
     0 &     25 &  &       3165.482 &       0.243 &  &       3165.537 &       0.231 &     1.91 &  &       3165.555 &       0.233 &  &       3165.580 &       0.234 &     1.60 \\[0.45ex]
     1 &     25 &  &       3223.052 &       0.226 &  &       3223.183 &       0.222 &     2.23 &  &       3223.302 &       0.246 &  &       3223.279 &       0.246 &     1.74 \\[0.45ex]
     2 &     25 &  &       3275.789 &       0.587 &  &       3276.510 &       0.536 &     0.67 &  &       3275.749 &       0.585 &  &       3275.754 &       0.545 &     0.53 \\[0.45ex]
     0 &     26 &  &       3286.718 &       0.298 &  &       3286.792 &       0.306 &     1.25 &  &       3286.935 &       0.365 &  &       3287.007 &       0.329 &     0.97 \\[0.45ex]
     1 &     26 &  &       3344.137 &       0.294 &  &       3344.679 &       0.287 &     1.50 &  &       3344.479 &       0.386 &  &       3344.531 &       0.363 &     1.11 \\[0.45ex]
     2 &     26 &  &       3397.777 &       0.684 &  &       3398.078 &       0.608 &     0.46 &  &       3397.302 &       0.668 &  &       3397.091 &       0.721 &     0.34 \\[0.45ex]
     0 &     27 &  &       3408.187 &       0.477 &  &       3409.214 &       0.433 &     0.81 &  &       3408.663 &       0.512 &  &       3408.606 &       0.532 &     0.61 \\[0.45ex]
     1 &     27 &  &       3465.693 &       0.433 &  &       3466.067 &       0.419 &     0.93 &  &       3466.404 &       0.472 &  &       3466.430 &       0.499 &     0.69 \\[0.45ex]
     2 &     27 &  &       3518.572 &       0.951 &  &       3518.691 &       1.077 &     0.26 &  &       3520.401 &       1.085 &  &       3519.568 &       1.151 &     0.21 \\[0.45ex]
     0 &     28 &  &       3531.333 &       0.741 &  &       3531.243 &       0.775 &     0.41 &  &       3531.755 &       0.927 &  &       3532.584 &       1.107 &     0.34 \\[0.45ex]
     1 &     28 &  &       3587.270 &       0.657 &  &       3587.792 &       0.810 &     0.53 &  &       3587.318 &       1.097 &  &       3587.141 &       1.312 &     0.39 \\[0.45ex]
     2 &     28 &  &       3640.416 &       1.853 &  &       3641.615 &       1.818 &     0.17 &  &       3641.946 &       1.253 &  &       3643.727 &       2.903 &     0.12 \\[0.45ex]
     0 &     29 &  &       3651.161 &       0.840 &  &       3650.143 &       1.027 &     0.28 &  &       3650.679 &       0.795 &  &       3649.769 &       1.716 &     0.19 \\[0.45ex]
     1 &     29 &  &       3710.839 &       0.840 &  &       3710.035 &       0.897 &     0.36 &  &       3710.678 &       0.938 &  &       3710.382 &       1.182 &     0.27 \\[0.45ex]
     2 &     29 &  &       3762.282 &       1.692 &  &       3771.014 &       7.568 &     0.12 &  &       3762.746 &       1.892 &  &       3764.758 &       2.095 &     0.10 \\[0.45ex]
     0 &     30 &  &       3769.722 &       1.096 &  &       3768.325 &       3.623 &     0.19 &  &       3770.309 &       1.537 &  &       3769.174 &       1.213 &     0.19 \\[0.45ex]
     1 &     30 &  &       3836.226 &       1.273 &  &       3837.493 &       1.676 &     0.20 &  &       3835.787 &       1.141 &  &       3836.531 &       0.888 &     0.24 \\[0.45ex]
 \hline 
\end{tabular}
\vskip-5pt
 \end{table*}
\end{appendix}
\end{document}